\documentclass[12pt]{article}
\usepackage[dvips]{color}
\usepackage{epsfig}
\usepackage{amsmath}
\usepackage{cite}
\usepackage{color}
\usepackage{subfigure}

\begin{document}
\begin{center}
\Large{\bf Exponential corrected thermodynamics of black holes}\\
\small \vspace{1cm}
{\bf Behnam Pourhassan$^{b}$\footnote {Email:~~~b.pourhassan@du.ac.ir}}
\\
\vspace{0.5cm}$^{b}${School of Physics, Damghan University, Damghan, 3671641167, Iran.}

\small \vspace{1cm}
\end{center}
\begin{abstract}
Recently, it is reported that thermal fluctuations, which are interpreted as quantum effects, modify the black holes entropy by an exponential term. We now find the effect of such modification on the black hole mass and other thermodynamics quantities. We find that the Schwarzschild black hole mass decreased by thermal fluctuations. Hence, we study exponential corrected thermodynamics and statistics of black holes by computing the partition function. We obtain the special condition on the event horizon radius to satisfy Smarr-Gibbs-Duhem relation in the presence of quantum correction. As we know, the Schwarzschild black hole is unstable, while the effect of exponential correction is the stability of $4D$ Schwarzschild black hole as well as the Schwarzschild-AdS black hole at a small area. On the other hand, a $5D$ Schwarzschild black hole is completely unstable. The effect of the quantum correction on the Reissner-Nordstr\"{o}m black hole is instability at quantum scales. Finally, we consider the most general case of charged AdS black hole and study the corrected thermodynamics.\\\\

\end{abstract}
\newpage

\section{Introduction}
Thermodynamics and related topics are important ways to study the black hole physics. Holographic principle help us to relate the surface quantities to the black hole thermodynamics \cite{BH+H}. Black hole entropy and temperature holographically related to the horizon area and surface gravity respectively \cite{1,2}. Black hole parameters like mass, charges, or angular momenta are thermodynamics variables of the first law which yields to the Smarr formula \cite{smarr}. In that case, generalization of the Smarr formula for both static and rotating black holes with nonlinear electromagnetic
fields has been proven and analyzed by Ref. \cite{1710.04660}. According to the quantum mechanics, a black hole can emit radiation which is known as Hawking radiation \cite{Ha1, Ha2}. Due to the Hawking radiation, a black hole decreases its size which may yield to evaporation \cite{eva001}. Also, a black hole may stop evaporation and become stable at the quantum ground state. From the string theory point of view, it is described by a space-time metric whose gauge fields denoted by mass, charges, or angular momenta \cite{eva1}. One of the best ways to investigate what happened to the black hole at the quantum scale is the study of thermal fluctuations \cite{ther1, ther2, ther3}. It yields to knowledge about the microscopic origin of entropy \cite{VS}. Indeed, it is the statistical fluctuations that may be interpreted as the quantum correction \cite{upadplb}. The study of such quantum corrections in a strong gravitational system needs a theory of quantum gravity. It can be done via some quantum theories of gravity like string theory or loop quantum gravity.\\
Already, it is found that leading order correction to the black hole entropy may be logarithmic \cite{log3, log4, log5, log6, log66, log7}. It is indeed an important term when the black hole size is small. Hence, it can be considered to test the quantum gravity \cite{test001, test002, test003, test0033, test004}. Such thermal fluctuations are considered as small perturbations around the equilibrium temperature. In that case, the effects of logarithmic correction on a BTZ black hole \cite{BTZ0} in a massive gravity investigated to find that thermal fluctuations modify the black hole stability \cite{BTZ1, BTZ10, BTZ2}. Logarithmic corrected (leading order) thermodynamics of Horava-Lifshitz black hole also investigated \cite{NPB2}, while the extension to the higher order correction is including inverse of entropy \cite{EPJC003}. The higher order corrections is indeed a way to calculate the microcanonical black hole entropy \cite{higher001}. It can be applied to the Schwarzschild and BTZ black Holes \cite{higher002}. It has been argued that corrections to the black hole entropy have universal form but different quantum gravity theories yield to different correction coefficients.\\
The logarithmic and the second order corrections are indeed perturbative corrections which are important when the black hole size reduced due to the Hawking radiation. Decreasing more we need non-perturbative analysis which is the main issue of this paper.
Recently, an exponential term proposed to correct the black hole entropy \cite{2007.15401}. It has been claimed that the exponential corrections in the black hole entropy may arise in any quantum theory of gravity \cite{2007.15401}. This term is negligible for a large horizon radius, while it is important when the black hole seizes become small. Until now, there is no thermodynamics analysis of black holes with exponential entropy corrected expect in \cite{CQG000}, hence it has been done in this paper for some kinds of important black holes.\\
As we know, any thermodynamics systems consist of some thermal fluctuations which yields to the ordinary entropy plus a logarithmic term as perturbative correction. However, there are a contribution of non-perturbative correction. Since any given thermodynamics system like black holes can including both corrections, there is an extra term in the entropy. In this paper, we show that the non-perturbative correction yields an exponential term as discussed by \cite{2007.15401}. Then, we try to obtain the effect of this correction on the partition function, hence study black hole exponential corrected thermodynamics.\\
This paper is organized as follows, in next section, we introduce exponential correction on the entropy. We use it in section \ref{thermo} to obtain partition function and hence study thermodynamics quantities. In section \ref{Sch}, we consider Schwarzschild black hole to apply the obtained formula of section \ref{thermo}. In section \ref{RN}, we consider Reissner-Nordstr\"{o}m black hole and find the effect of exponential term on the black hole stability. In section \ref{SchAdS}, we extend our calculations to the case of the Schwarzschild-AdS black hole. The more general case of charged AdS black hole discussed in section \ref{Ch}. Finally, in section \ref{Con}, we give a conclusion and summary of results.

\section{Corrections on black hole entropy}
As we know, the entropy of large black holes (comparing with the Planck scale) is proportional to the horizon area. Also, we know that the black hole size is reduced due to the Hawking radiation. Hence, the small black hole entropy needs some corrections. These corrections may interpreted as a quantum effect, coming from thermal fluctuations, which yields to the modification of the holographic principle \cite{H1, H2}. It has been found that the leading order corrections to the Bekenstein-Hawking entropy is logarithmic \cite{log0}. However, for the space-time dimension other than four ($D\neq4$), the black hole entropy corrections may be power law \cite{P1} which is like higher order corrections \cite{higher3}. We should note that the entropy corrected black holes may be investigated by using the non-perturbative quantum theory of general relativity. In that case, the leading order entropy corrected AdS black hole in the large area limit of a four-dimensional Einstein gravity with negative cosmological constant has been studied by Ref. \cite{18}. Moreover, it is interesting to consider a matter field near the extremal Reissner-Nordstrom and dilatonic black holes and study the effect of quantum corrections on the black hole thermodynamics \cite{19}. Logarithmic corrected thermodynamics is an interesting subject in various contexts like that discussed by Refs. \cite{22, 23}. Several black objects are considered to investigate universality in the form of correction terms \cite{26, 27}. It is also possible to use the partition function to study the corrected thermodynamics of black hole \cite{29}, which yields to the fact that the quantum correction to space-time structure would produce thermal fluctuations \cite{32}.\\
Assuming a black hole consists of total $N$ particles we can write total microstates to obtain the entropy. From the statistical mechanics we know that, total number of microstates of a given system expressed as
\begin{equation}\label{Sm}
\Omega=\frac{\left(\Sigma{n_{i}}\right)!}{\Pi{n_{i}}!},
\end{equation}
Assuming each number $n_{i}$ is shared by $s_{i}$ pieces, therefore,
\begin{equation}\label{Sm1}
\Omega=\frac{\left(\Sigma{s_{i}}\right)!}{\Pi{s_{i}}!}.
\end{equation}
Also, $\varepsilon_{i}$ is the energy of the $i$th microstate with number $n_{i}$. Hence
\begin{equation}\label{St1}
N=\Sigma{s_{i}n_{i}},
\end{equation}
is total number and
\begin{equation}\label{St2}
E=\Sigma{s_{i}n_{i}\varepsilon_{i}},
\end{equation}
is total energy. Then, using the Stirling formula for the large $N$ limit,
\begin{equation}\label{SmO}
\ln{N!}=N\ln{N}-N,
\end{equation}
and varying $\ln{\Omega}$ under the following conditions,
\begin{eqnarray}\label{S000}
\delta\Sigma{s_{i}n_{i}}&=&0,\nonumber\\
\Sigma{\varepsilon_{i}\delta(s_{i}n_{i})}&=&0,
\end{eqnarray}
one can obtain the most likely configuration as
\begin{equation}\label{S00}
s_{i}=\left(\sum{n_{i}}\right)e^{-\lambda n_{i}},
\end{equation}
where $\lambda$ is called the variation parameter (plays role of Lagrange multipliers) which satisfied the following condition \cite{2007.15401},
\begin{equation}\label{lambda}
\sum{e^{-\lambda n_{i}}}=1,
\end{equation}
which yields,
\begin{equation}\label{lambda1}
\lambda\approx\ln{2}-2^{-N},
\end{equation}
where $\mathcal{O}(2^{-2N})$ neglected. So, the entropy given by $S=\lambda N$.
Hence, eliminating $N$ using the equation (\ref{lambda1}) tells that the quantum correction to the black hole entropy is exponential and given by,
\begin{equation}\label{S}
S=S_{0}+e^{-S_{0}},
\end{equation}
with
\begin{equation}\label{S0}
S_{0}=\frac{A}{4 l_{p}^{2}},
\end{equation}
where $l_{p}$ is Planck length and $A$ is the black hole horizon area.\\
All above mentioned corrections can be expressed as following,
\begin{equation}\label{S-total}
S=S_{0}+\alpha \ln{f_{1}(S_{0})}+\frac{\gamma}{S_{0}}+ \eta e^{f_{2}(S_{0})},
\end{equation}
where $\alpha$, $\gamma$ and $\eta$ are some infinitesimal constants which are called correction coefficients. Also, $f_{1}(S_{0})$ and $f_{2}(S_{0})$ are suitable functions of the uncorrected black hole entropy.
Several forms of function $f_{1}(S_{0})$ like $S_{0}T^{2}$, $CT^{2}$ or $S_{0}$ already discussed in literatures \cite{f1, f2, f3, f4, f5}. The logarithmic correction with coefficient $\alpha$ \cite{log1, log2} and higher order correction with coefficient $\gamma$ \cite{higher1, higher2} may be negligible. Hence, it is possible to consider the case $\alpha=\gamma=0$ which is interested in this paper.
In Ref. \cite{2007.15401}, it is argued that $f_{2}(S_{0})=-S_{0}$ and $\eta=1$, independent of the theory of quantum gravity. The exponential term also is negligible for the large black hole areas (entropy), but its effect is important when the black hole area is small, Therefore, it is considered as a quantum effect for the small black hole as well as logarithmic correction \cite{NPB}. It should be noted that the exponential correction on the statistical entropy of supersymmetric string theories using the quantum entropy function formalism \cite{Sen001}. It is compatible with the non-perturbative features of the string theory \cite{string} and is valid in the Planckian regime of the black hole event horizon area. It is clear that for the large black hole we have $A\gg1$ so $S_{0}\gg1$, hence $S=S_{0}$. On the other hand, for the case of $S_{0}=0$ ($A=0$), we have $S=1$, so we may use it for the two-dimensional black holes as well as entropy function formalism \cite{e1, e2}. Although, quantum entropy function may also yield to the logarithmic correction \cite{e3}. In general, according to the relation (\ref{S}) we can find $S\geq1$.\\
A general spherically symmetric black hole metric in $D$ dimensional space-time given by,
\begin{equation}\label{metric}
ds^{2}=-f(r)dt^{2}+\frac{dr^{2}}{f(r)}+r^{2}d\Omega_{D-2}^{2}.
\end{equation}
Black hole area, and hence, black hole entropy depends on $d\Omega_{D-2}^{2}$ elements, while black hole temperature (at outer horizon radius $r_{+}$) is given by,
\begin{equation}\label{temp}
T=\frac{1}{4\pi}\left(\frac{df(r)}{dr}\right)_{r=r_{+}},
\end{equation}
which is independent of $d\Omega_{D-2}^{2}$ elements. Therefore, the black hole entropy may corrected but the black hole temperature remains unchanged.
However, there are some possible ways to obtain corrected temperature \cite{CT}. In this work, we assume that the Hawking temperature of the black hole given by the equation (\ref{temp}) do not affect by quantum corrections. Now, we justify these assumption by proposing the following line element,
\begin{equation}\label{metric2}
ds^{\prime2}=-f(r)dt^{2}+\frac{dr^{2}}{f(r)}+r^{2}K(r)d\Omega_{D-2}^{2},
\end{equation}
where
\begin{eqnarray}\label{metric2function}
K(r)&=&1+\alpha\frac{4l_{p}^{2}}{r_{H}^{D-2}\Omega_{D-2}}\ln{f_{1}\left(\frac{r_{H}^{D-2}\Omega_{D-2}}{4l_{p}^{2}}\right)}\nonumber\\
&+&\gamma\frac{(4l_{p}^{2})^{2}}{(r_{H}^{D-2}\Omega_{D-2})^{2}}+
\eta\frac{4l_{p}^{2}}{r_{H}^{D-2}\Omega_{D-2}}\exp{f_{2}\left(\frac{r_{H}^{D-2}\Omega_{D-2}}{4l_{p}^{2}}\right)}+...,
\end{eqnarray}
where dots represent higher order terms of correction coefficients which neglected. So, the modified black hole horizon area is given by,
\begin{equation}\label{area2}
\mathcal{A}=K(r_{+})r_{+}^{D-2}\Omega_{D-2},
\end{equation}
hence the corrected entropy (\ref{S-total}) reproduced by,
\begin{equation}\label{S}
S=\frac{\mathcal{A}}{4 l_{p}^{2}}.
\end{equation}
Using the metric (\ref{metric2}), it is clear that the event horizon radius remains unchanged and the black hole temperature, as before, given by the equation (\ref{temp}). Because the additional terms of (\ref{metric2function}) are small perturbations around the equilibrium, hence have not any effective impact on the field equations and hence the geometry of $d\Omega^{2}$. It is like the situation happen for the Ho\v{r}ava-Lifshitz black hole \cite{EPJC003}. This is also true for the corrections obtained using AdS/CFT \cite{adscft}.\\
The extra exponential term of the black hole entropy (\ref{S}) affects some thermodynamics quantities which will be discussed in next section. Also, it modifies the partition function of statistical physics, which will obtained in next section.
\section{Thermodynamics and statistics}\label{thermo}
\begin{figure}[h!]
 \begin{center}$
 \begin{array}{cccc}
\includegraphics[width=70  mm]{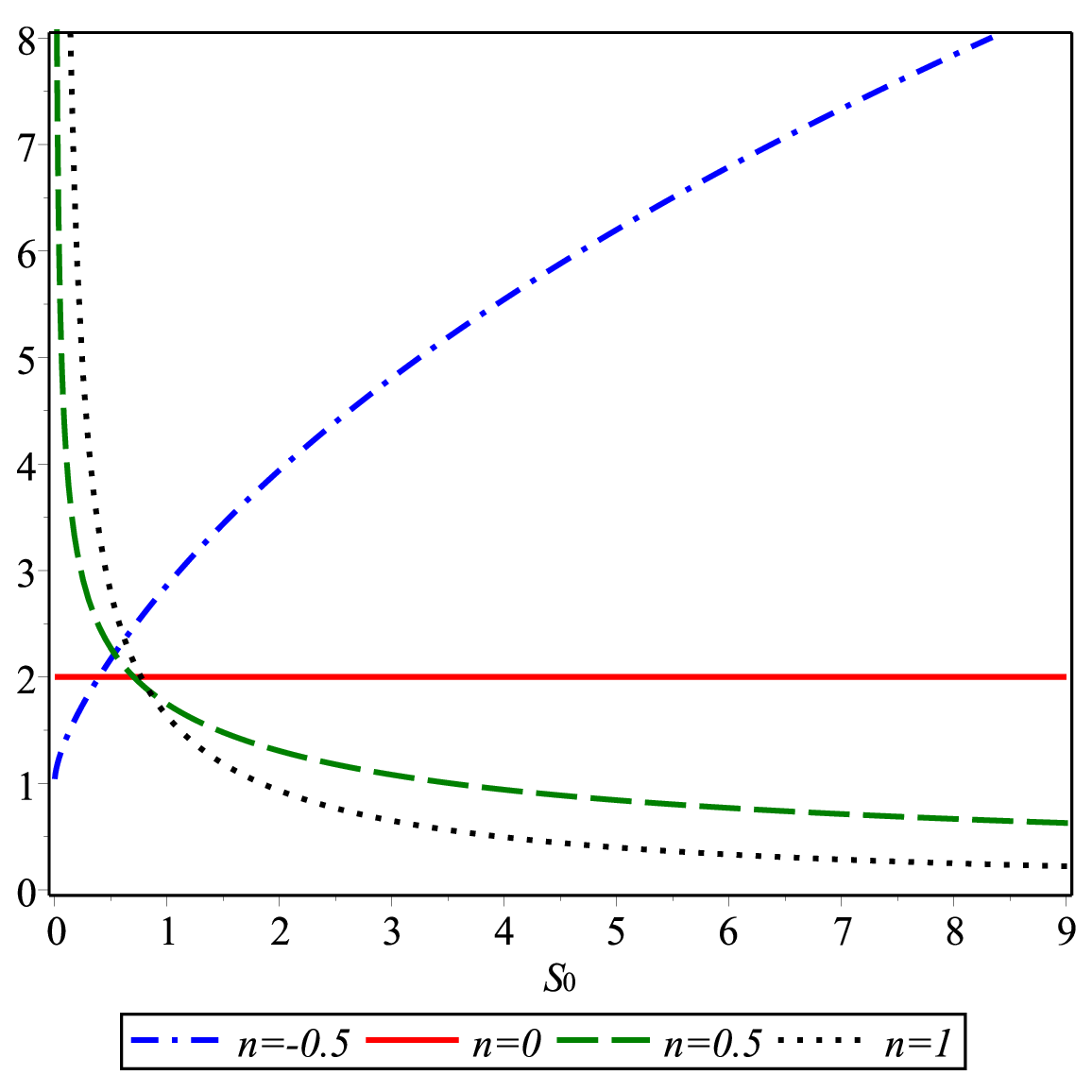}
 \end{array}$
 \end{center}
\caption{$g_{q}(S_{0})$ in terms of $S_0$ for $g_{q}=1$.}
 \label{fig1}
\end{figure}

In this section, we would like to use corrected entropy (\ref{S}) to see a modification of some thermodynamics and statistics quantities in a general form. To find a general formalism, we need to specify the temperature dependence of the entropy. It is a power law for several systems like Schwarzschild black holes. Hence, we use the following ansatz,
\begin{equation}\label{T}
T=c_{1}S_{0}^{n},
\end{equation}
where $c_{1}$ is a constant and $n$ is a number. Also, it is easy to extend this relation as,
\begin{equation}\label{TT}
T=\sum_{i}c_{i}S_{0}^{n_{i}}.
\end{equation}
Hence, the equation (\ref{T}) is a special case of the equation (\ref{TT}) where there is only one coefficient $c_{1}$.\\
To begin, we use a well-known relation between entropy and partition function ($Z$) of a canonical
ensemble which is,
\begin{equation}\label{SZ}
S=\ln{Z}+T\left(\frac{\partial \ln{Z}}{\partial T}\right)_{V}.
\end{equation}
By using the exponential corrected entropy (\ref{S}) and temperature (\ref{T}) in the relation (\ref{SZ}) we obtain the following relation,
\begin{equation}\label{LnZ}
\ln{Z}=\frac{n}{n+1}S_{0}+\frac{n}{S_{0}^{n}}\int{S_{0}^{n-1}e^{-S_{0}}dS_{0}}+g(S_{0}),
\end{equation}
where $g(S_{0})$ is an unknown function depend on quantum correction which will be obtained using other thermodynamics quantities. Indeed, quantum effects are encoded in both the last terms of the expression (\ref{LnZ}). Hence, at the first step, we try to solve the above integral numerically. By using the following relation,
\begin{equation}\label{int}
\int_{0}^{\infty}{x^{n}e^{-x}dx}=n!,
\end{equation}
we can rewrite the equation (\ref{LnZ}) as following,
\begin{equation}\label{LnZ1}
\ln{Z}=\frac{n}{n+1}S_{0}+\frac{n(n-1)!}{S_{0}^{n}}+g(S_{0}).
\end{equation}
The function $g(S_{0})$ includes two parts, one is corresponding to ordinary entropy (classical) and the other is corresponding to quantum correction, hence we can write,
\begin{equation}\label{g}
g(S_{0})\equiv\frac{g_{c}}{S_{0}^{n}}+g_{q}(S_{0}),
\end{equation}
where $g_{c}$ is a constant related to the classical part, while $g_{q}(S_{0})$ is an unknown function due to the quantum correction. This part added to recover the missed entropy-dependence part due to the numerical calculations of (\ref{int}). Therefore,
\begin{equation}\label{LnZ2}
\ln{Z}=\ln{Z_{0}}+\ln{Z_{c}}=\left[\frac{n}{n+1}S_{0}+\frac{g_{c}}{S_{0}^{n}}\right]+\left[\frac{n(n-1)!}{S_{0}^{n}}+g_{q}(S_{0})\right],
\end{equation}
where
\begin{equation}\label{LnZc}
\ln{Z_{c}}=\frac{n(n-1)!}{S_{0}^{n}}+g_{q}(S_{0}),
\end{equation}
referred to the correction of partition function due to the thermal fluctuations, while $Z_0$ denotes uncorrected partition function. Therefore, the corrected partition function can express as,
\begin{equation}\label{Z}
Z=\exp\left(\frac{n}{n+1}S_{0}+\frac{g_{c}+n(n-1)!}{S_{0}^{n}}+g_{q}(S_{0})\right).
\end{equation}
Having the partition function, help us to obtain all thermodynamics quantities. For example, internal energy given by,
\begin{eqnarray}\label{U}
U&=&T^{2}\frac{d\ln{Z}}{dT}\nonumber\\
&=&\frac{c_{1}}{n}S_{0}^{n+1}\left[\frac{n}{n+1}-\frac{n}{S_{0}^{n+1}}(g_{c}+n(n-1)!)+g_{q}^{\prime}(S_{0})\right],
\end{eqnarray}
where prime denotes derivatives with respect to $S_{0}$.\\
On the other hand, having corrected entropy and temperature also gives us the thermodynamics of a given system. Both ways should coincide and this helps us to determine $g_{q}(S_{0})$ function. We can do that by using the Helmholtz free energy $F$. According to the statistical mechanical we have,
\begin{equation}\label{Fs}
F=-T\ln{Z}=-c_{1}S_{0}^{n}\left[\frac{n}{n+1}S_{0}+\frac{g_{c}+n(n-1)!}{S_{0}^{n}}+g_{q}(S_{0})\right],
\end{equation}
while according to the thermodynamics we have,
\begin{equation}\label{Ft}
F=U-TS=U-T(S_{0}+e^{-S_{0}}).
\end{equation}
Both equations (\ref{Fs}) and (\ref{Ft}) should make the same result. It yields to the following differential equation,
\begin{equation}\label{diff1}
S_{0}g_{q}^{\prime}(S_{0})+ng_{q}(S_{0})=ne^{-S_{0}}.
\end{equation}
Solution of the above equation obtained as,
\begin{equation}\label{sol1}
g_{q}(S_{0})=\frac{g_{q}}{S_{0}^{n}}+\left[\frac{WM(\frac{n}{2},\frac{n+1}{2},S_{0})S_{0}^{-\frac{n}{2}}}{n+1}
+WM(\frac{n+2}{2},\frac{n+1}{2},S_{0})S_{0}^{-\frac{n+2}{2}}\right]e^{-\frac{S_{0}}{2}},
\end{equation}
where $WM(\mu, \nu, z)$ is Whittaker function which is constructed from hypergeometric function. Also, $g_{q}$ is an integration constant. In the case of $n=0$ we have $g_{q}(S_{0})\approx1+g_{q}$. In Fig. \ref{fig1} we can see behavior of $g_{q}(S_{0})$ in terms of $S_0$ for some values of $n$. Hence, using the solution (\ref{sol1}) in the equation (\ref{LnZ2}) one can write,
\begin{eqnarray}\label{LnZ3}
\ln{Z}&=&\frac{n}{n+1}S_{0}+\frac{g+n(n-1)!}{S_{0}^{n}}\nonumber\\
&+&\left[\frac{WM(\frac{n}{2},\frac{n+1}{2},S_{0})S_{0}^{-\frac{n}{2}}}{n+1}
+WM(\frac{n+2}{2},\frac{n+1}{2},S_{0})S_{0}^{-\frac{n+2}{2}}\right]e^{-\frac{S_{0}}{2}},
\end{eqnarray}
where $g\equiv g_{c}+g_{q}$ is used.\\

\begin{figure}[h!]
 \begin{center}$
 \begin{array}{cccc}
\includegraphics[width=55 mm]{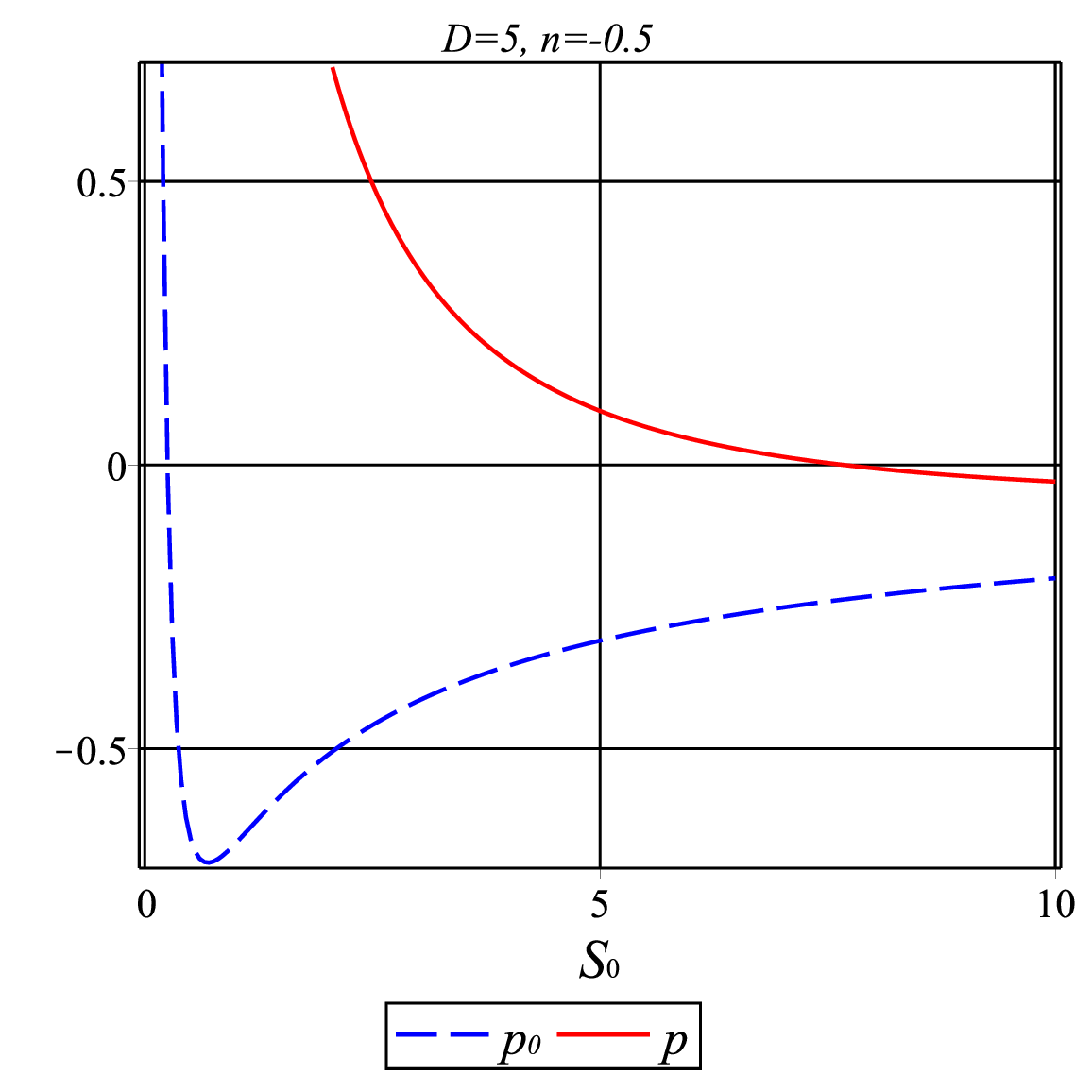}\includegraphics[width=55 mm]{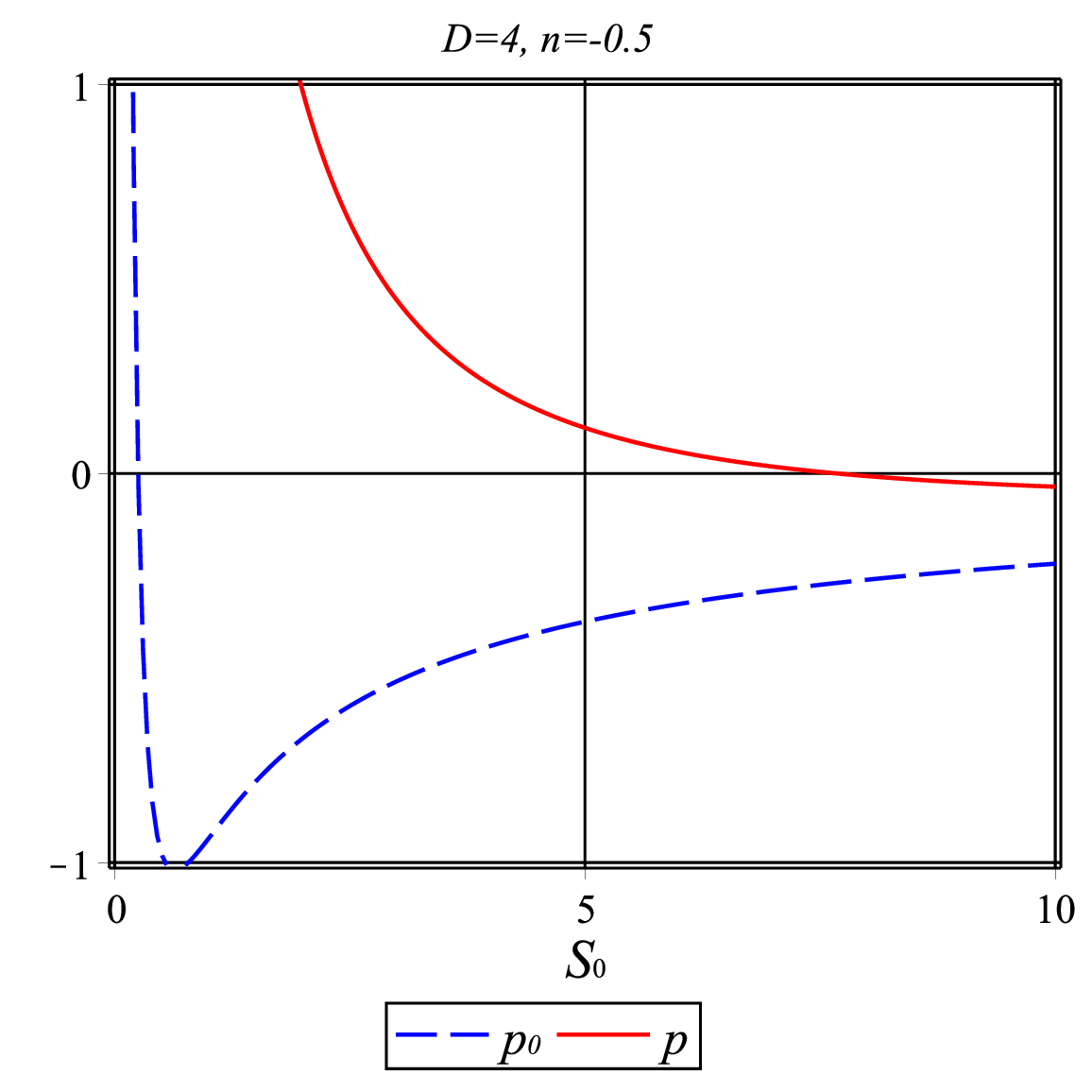}
 \end{array}$
 \end{center}
\caption{Pressure in terms of $S_0$ for $g=1$.}
 \label{fig2}
\end{figure}

To investigate further about thermodynamics quantities like pressure, enthalpy, and Gibbs free energy we need black hole volume, which can be expressed as following,
\begin{equation}\label{V}
V=v_{d}S_{0}^{\frac{d}{d-1}},
\end{equation}
where $v_{d}$ is a constant and $d$ is spatial dimension. For example, in five-dimensional space-time ($d=4$) we find $v_{d}=(\frac{2}{\pi^{2}})^{1/3}$.\\
In that case, pressure given by,
\begin{equation}\label{p}
p=T\frac{d\ln{Z}}{dV}=p_{0}+p_{c},
\end{equation}
where $p_{0}$ is uncorrected pressure given by,
\begin{equation}\label{p0}
p_{0}=\frac{(d-1)c_{1}n(S_{0}^{n+1}-g_{c}(n+1))}{(n+1)dv_{d}S_{0}^{\frac{d}{d-1}}},
\end{equation}
while $p_{c}$ is corrected pressure which is obtained as,
\begin{equation}\label{pc}
p_{c}=\frac{(d-1)c_{1}n}{(n+1)dv_{d}}\left[S_{0}^{\frac{n(d-1)-2d}{d-1}}e^{-\frac{S_{0}}{2}}WM(\frac{n}{2},\frac{n+1}{2},S_{0})
+S_{0}^{-\frac{d}{d-1}}\left((n+1)(n!+g_{q})\right)\right].
\end{equation}
In plots of Fig. \ref{fig2} we can see the effects of quantum correction on the black hole pressure. Increasing pressure is one of the important quantum effects.

\begin{figure}[h!]
 \begin{center}$
 \begin{array}{cccc}
\includegraphics[width=70 mm]{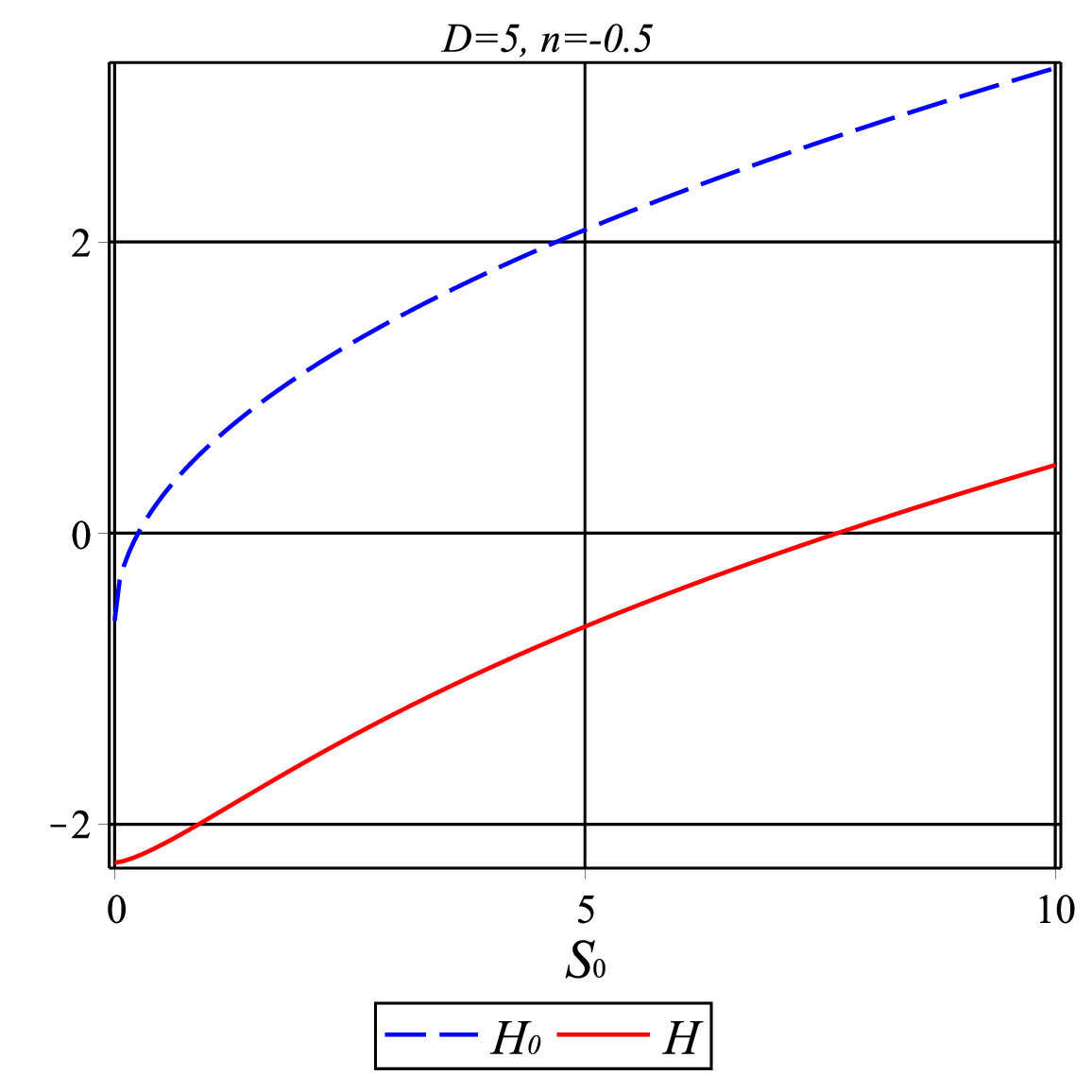}
 \end{array}$
 \end{center}
\caption{Enthalpy in terms of $S_0$ for $g=1$.}
 \label{fig3}
\end{figure}

Enthalpy is given by the following general formula,
\begin{equation}\label{H}
H=T\left[T\frac{d\ln{Z}}{dT}+V\frac{d\ln{Z}}{dV}\right]\equiv H_{0}+H_{c},
\end{equation}
where $H_{0}$ is uncorrected entropy, while $H_{c}$ denotes the effect of quantum correction which are respectively obtained as,
\begin{equation}\label{H0}
H_{0}=\frac{c_{1}((d-1)n+d)(S_{0}^{n+1}-g_{c}(n+1))}{(n+1)d},
\end{equation}
and
\begin{equation}\label{Hc}
H_{c}=-\frac{c_{1}((d-1)n+d)}{(n+1)d}\left[S_{0}^{\frac{n}{2}e^{-\frac{S_{0}}{2}}}WM(\frac{n}{2},\frac{n+1}{2},S_{0})+(n+1)!+g_{q}(n+1)\right].
\end{equation}
In Fig. \ref{fig3} we can see the behavior of the entropy and find the effect of quantum correction which is decreasing enthalpy.

\begin{figure}[h!]
 \begin{center}$
 \begin{array}{cccc}
\includegraphics[width=70 mm]{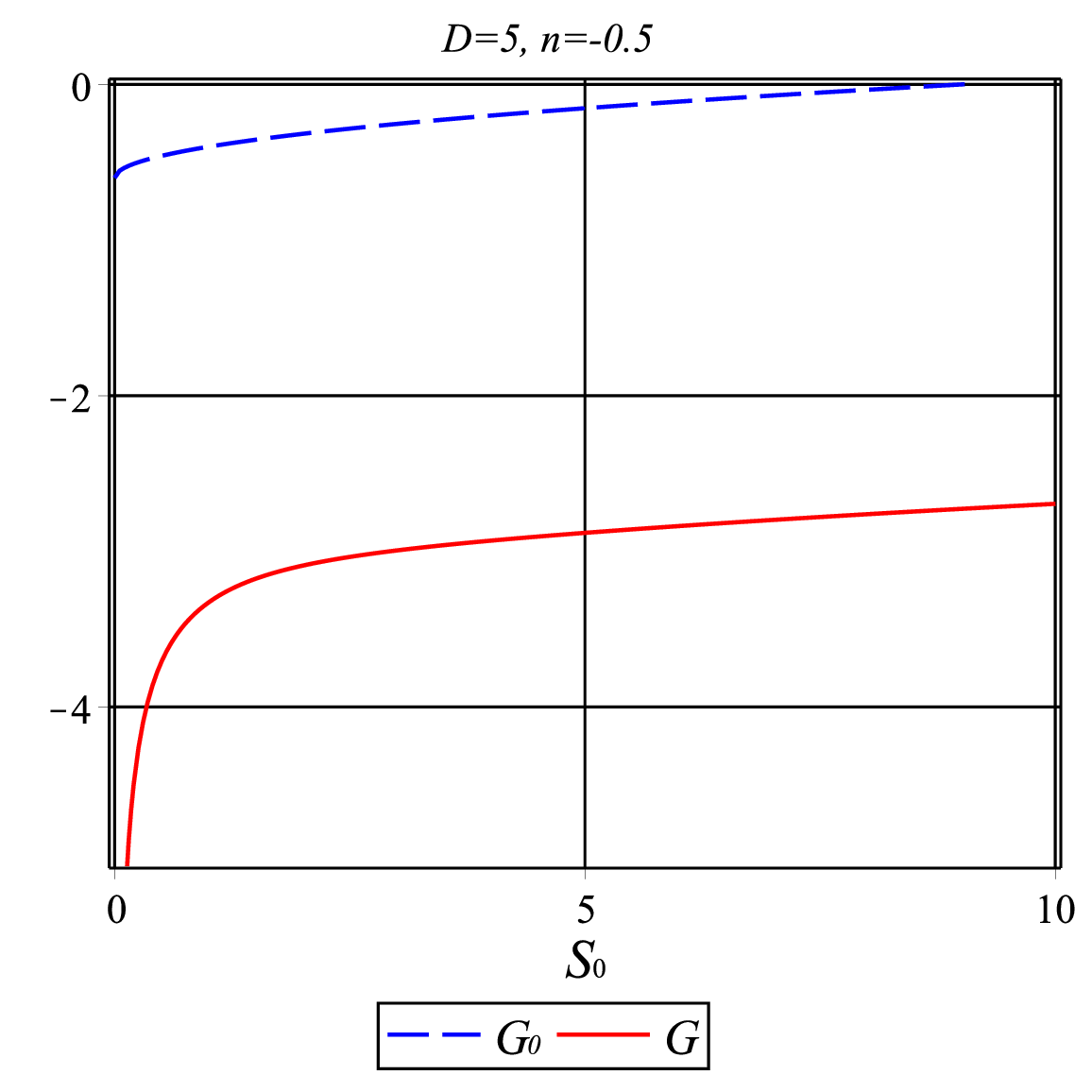}
 \end{array}$
 \end{center}
\caption{Gibbs free energy in terms of $S_0$ for $g=1$.}
 \label{fig4}
\end{figure}

Then, we can obtain Gibbs free energy via,
\begin{equation}\label{H}
G=T\left[-\ln{Z}+V\frac{d\ln{Z}}{dV}\right]\equiv G_{0}+G_{c}.
\end{equation}
In Fig. \ref{fig4} we can see the effect of quantum correction on the Gibbs free energy.

\begin{figure}[h!]
 \begin{center}$
 \begin{array}{cccc}
\includegraphics[width=55 mm]{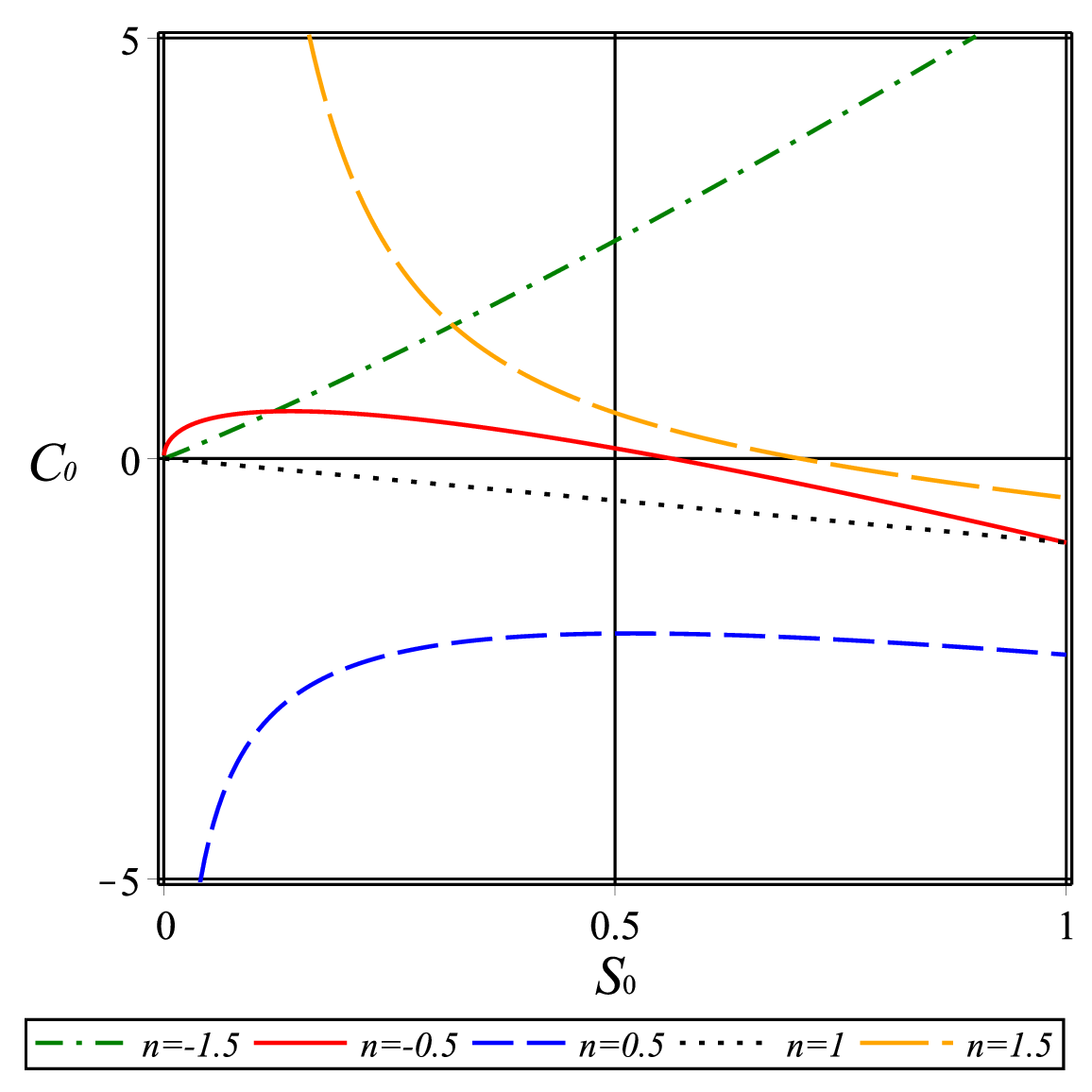}\includegraphics[width=55 mm]{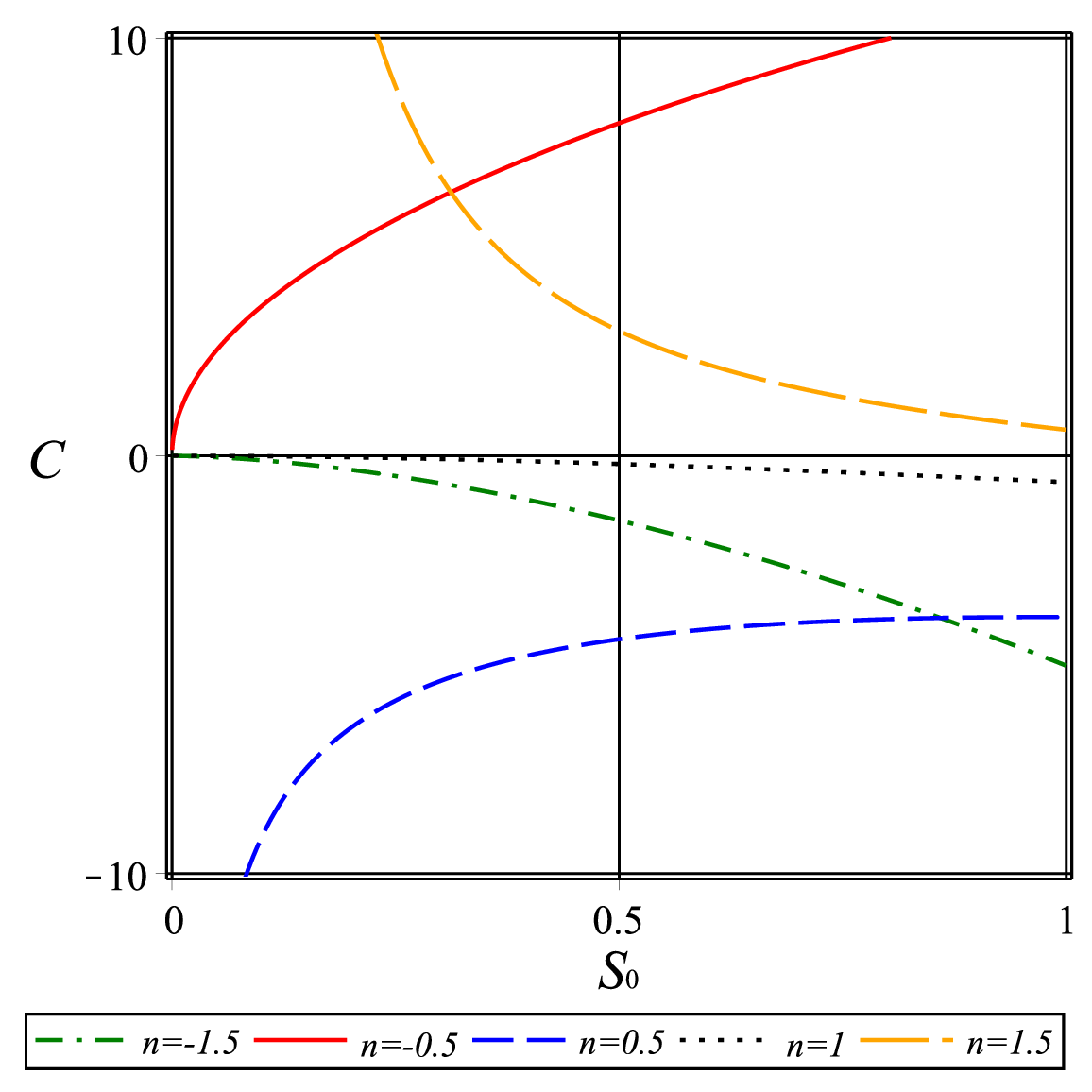}
 \end{array}$
 \end{center}
\caption{Specific heat in terms of $S_0$ for $g=1$.}
 \label{fig5}
\end{figure}

\begin{figure}[h!]
 \begin{center}$
 \begin{array}{cccc}
\includegraphics[width=70 mm]{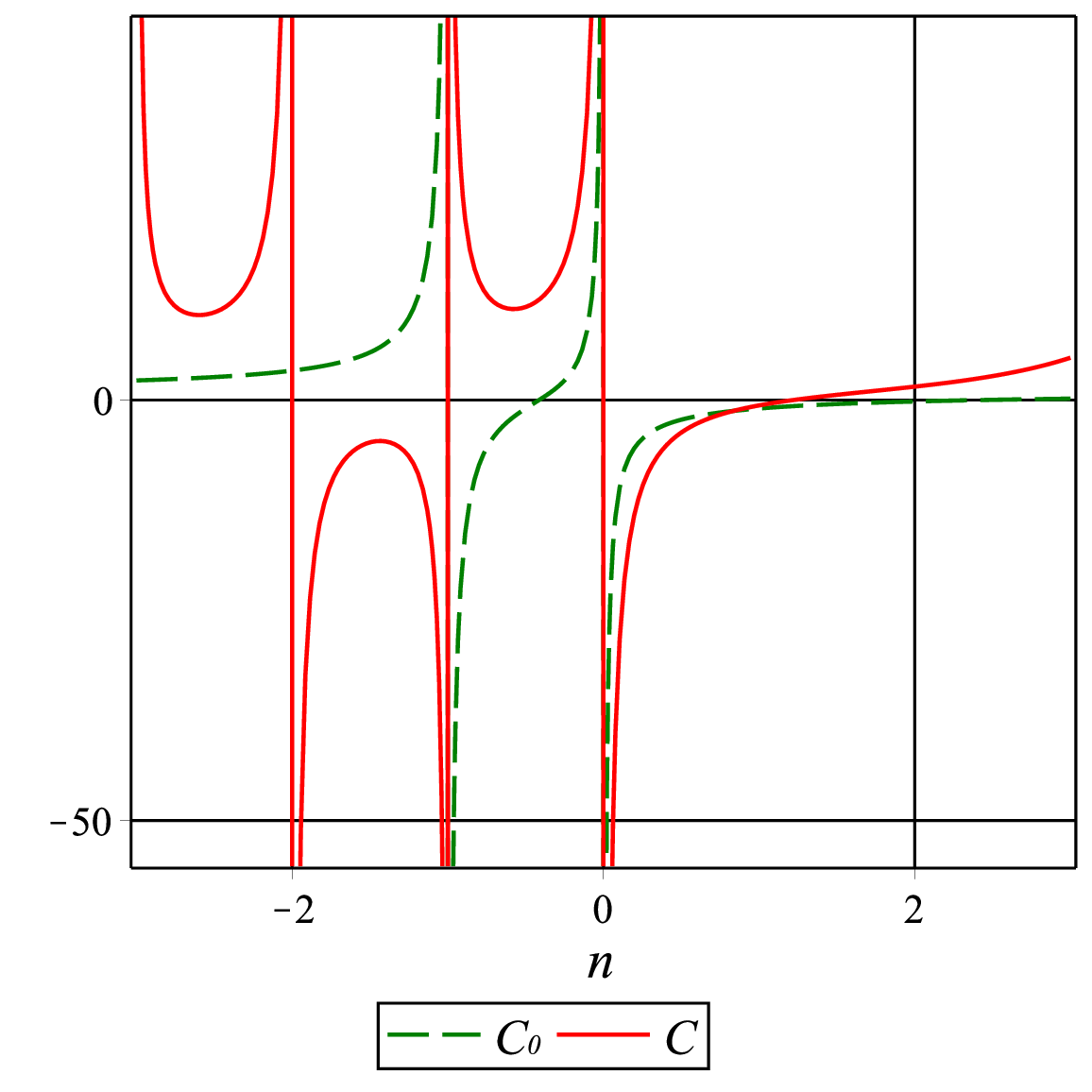}
 \end{array}$
 \end{center}
\caption{Specific heat in terms of $n$ for $S_{0}=g=1$.}
 \label{fig6}
\end{figure}

Finally, we can discuss specific heat at constant volume,
\begin{equation}\label{Cv}
C=T\left[2\frac{d\ln{Z}}{dT}+T\frac{d^{2}\ln{Z}}{dT^{2}}\right]\equiv C_{0}+C_{c},
\end{equation}
where
\begin{equation}\label{C0}
C_{0}=\frac{2nS_{0}^{n+1}-g_{c}(n-1)(n+1)}{n(n+1)S_{0}^{n}},
\end{equation}
and
\begin{equation}\label{Cc}
C_{c}=-\frac{S_{0}^{-\frac{n}{2}}e^{-\frac{S_{0}}{2}}WM(\frac{n}{2},\frac{n+1}{2},S_{0})+
(n+1)\left((n-1)(n\Gamma(n)+g_{q})S_{0}^{-n}+S_{0}e^{-S_{0}}\right)}{n(n+1)},
\end{equation}
In plots of Fig. \ref{fig5} we draw specific heat in terms of $S_{0}$ for some values of $n$. We can see that quantum correction is different for a given system with different $n$. For example, the system with $n=-1.5$ is stable ($C_{0}<0$) while in presence of a quantum effect, corrected specific heat is a negative and the black hole is unstable. In the case of $n=-0.5$ also the effect of quantum correction is the stability of the black hole (see the solid red line of Fig. \ref{fig5}). It means that some kind of black holes, reduced its size due to the Hawking radiation and goes to a stable phase hence do not evaporate.\\
Then, in the Fig. \ref{fig6}, we draw specific heat in terms of $n$ and see some phase transition for specific values of $n$ like $n=-1, -2$.\\
Now, we can consider some black hole kinds to exam about general formulation. We begin with the Schwarzschild black hole.

\section{Schwarzschild black hole}\label{Sch}
The simplest spherically symmetric solution is a Schwarzschild black hole. We consider it in both 4 and 5 dimensions separately to study exponential corrected thermodynamics. The black hole mass ($M$) is an only important parameter of this kind, hence the first law of thermodynamics for the Schwarzschild black hole given by,
\begin{equation}\label{First Law S}
dM=TdS=TdS_{0}(1-e^{-S_{0}}),
\end{equation}
where in the last equality we used (\ref{S}). Using the equation (\ref{T}) in the thermodynamics first law (\ref{First Law S}), then applying numerical integration (\ref{int}), we find,
\begin{equation}\label{MS}
M=c_{1}\left(\frac{S_{0}^{n+1}}{n+1}-n!\right)=M_{0}-c_{1}n!,
\end{equation}
which means quantum corrections reduced the value of the black hole mass. For obtaining expected results for usual Schwarzschild black hole thermodynamics, we should set $g=0$.\\
The Smarr-Gibbs-Duhem relation for the Schwarzschild black hole is given by \cite{1401.2586},
\begin{equation}\label{SGDS}
\frac{D-3}{D-2}M=TS.
\end{equation}
Following, we study two separate cases of Schwarzschild black hole in four and five dimensions.
\subsection{4D}
\begin{figure}[h!]
 \begin{center}$
 \begin{array}{cccc}
\includegraphics[width=70 mm]{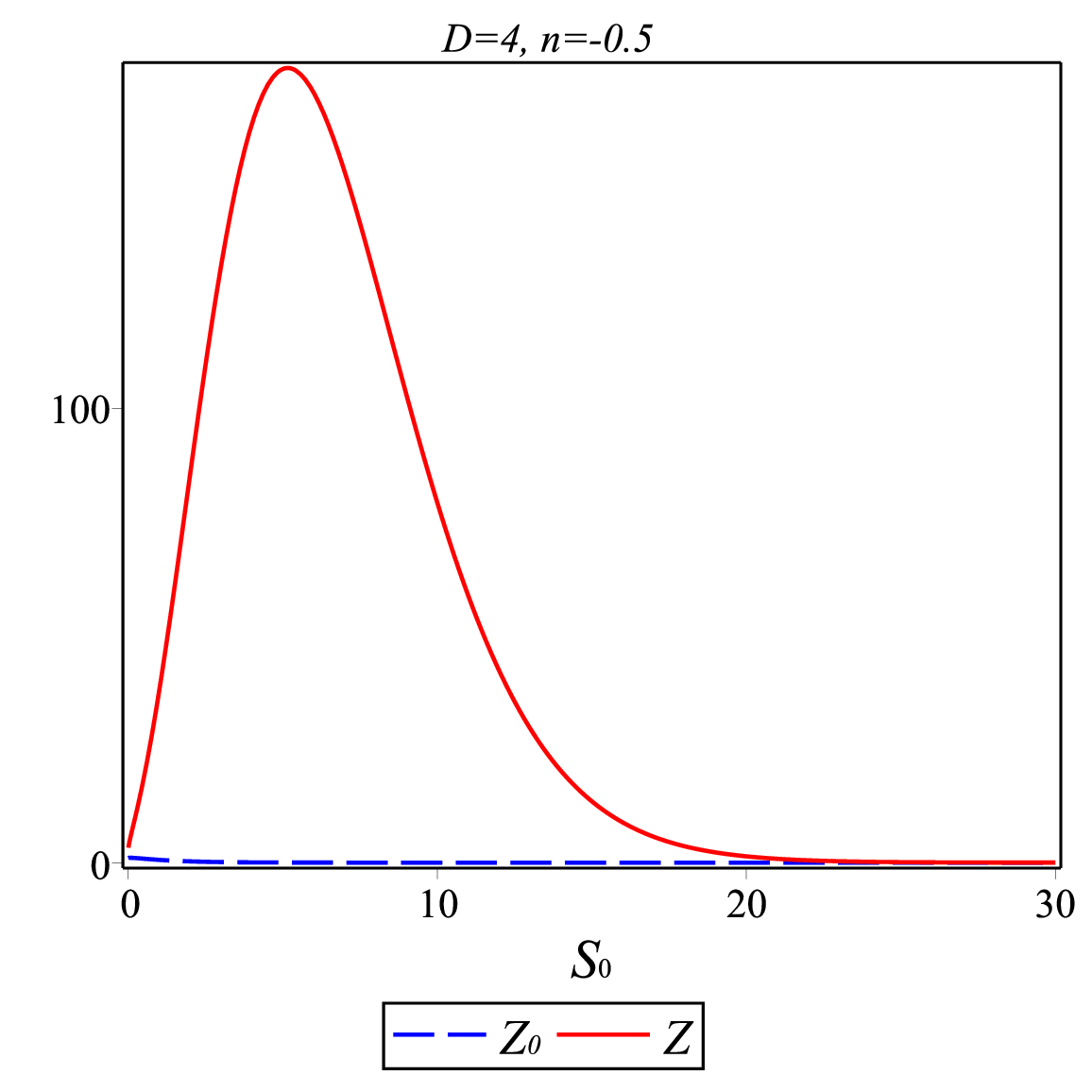}
 \end{array}$
 \end{center}
\caption{Partition function of $4D$ Schwarzschild black hole in terms of $S_{0}$ for $g=1$.}
 \label{fig7}
\end{figure}

Four-dimensional Schwarzschild black hole given by the metric (\ref{metric}) with $D=4$ and (see for example \cite{1401.2586}),
\begin{equation}\label{metricS4D}
f(r)=1-\frac{2M_{0}}{r},
\end{equation}
where $M_{0}=TdS_{0}$ denotes the uncorrected black hole mass satisfying ordinary first law of thermodynamics, so $r_{+}=2M_{0}$ is the black hole event horizon radius. The black hole entropy and thermodynamic volume given by,
\begin{equation}\label{SS4D}
S_{0}=\pi r_{+}^{2},
\end{equation}
and
\begin{equation}\label{VS4D}
V=\frac{4}{3}\pi r_{+}^{3},
\end{equation}
respectively.
Using the relations (\ref{temp}) and (\ref{metricS4D}) one can obtain,
\begin{equation}\label{TS4D}
T=\frac{1}{4\pi r_{+}}.
\end{equation}
Combination of (\ref{SS4D}) and (\ref{TS4D}) yields $T=\frac{1}{4\sqrt{\pi S_{0}}}$, hence we find $n=-\frac{1}{2}$ and $c_{1}=\frac{1}{4\sqrt{\pi}}$ in the equation (\ref{T}). Also, combination of (\ref{V}), (\ref{SS4D}) and (\ref{VS4D}) gives, $v_{d}=\frac{4}{3\pi}$, so $V=\frac{4}{3\pi}S_{0}^{\frac{3}{2}}$.\\
The first law of black hole thermodynamics reads as,
\begin{equation}\label{First Law S4D}
dM=dM_{0}(1-e^{-S_{0}}),
\end{equation}
which yield,
\begin{equation}\label{MS4D}
M=M_{0}-\frac{1}{4}.
\end{equation}
In that case, the Smarr-Gibbs-Duhem formula (\ref{SGDS}) gives
\begin{equation}\label{SGDS4D}
M_{0}-\frac{1}{4}=2T(S_{0}+e^{-S_{0}}),
\end{equation}
which hold only in the special radius $r_{+}\approx0.867$, where we used $M_{0}=2TS_{0}$ as ordinary (uncorrected) Smarr-Gibbs-Duhem relation.\\
In Fig. \ref{fig7} we can see the important effect of quantum correction on the partition function of $4D$ Schwarzschild black hole. Both solid and dashed lines of Fig. \ref{fig7} are coincide for large area (large $S_{0}$).

\begin{figure}[h!]
 \begin{center}$
 \begin{array}{cccc}
\includegraphics[width=70 mm]{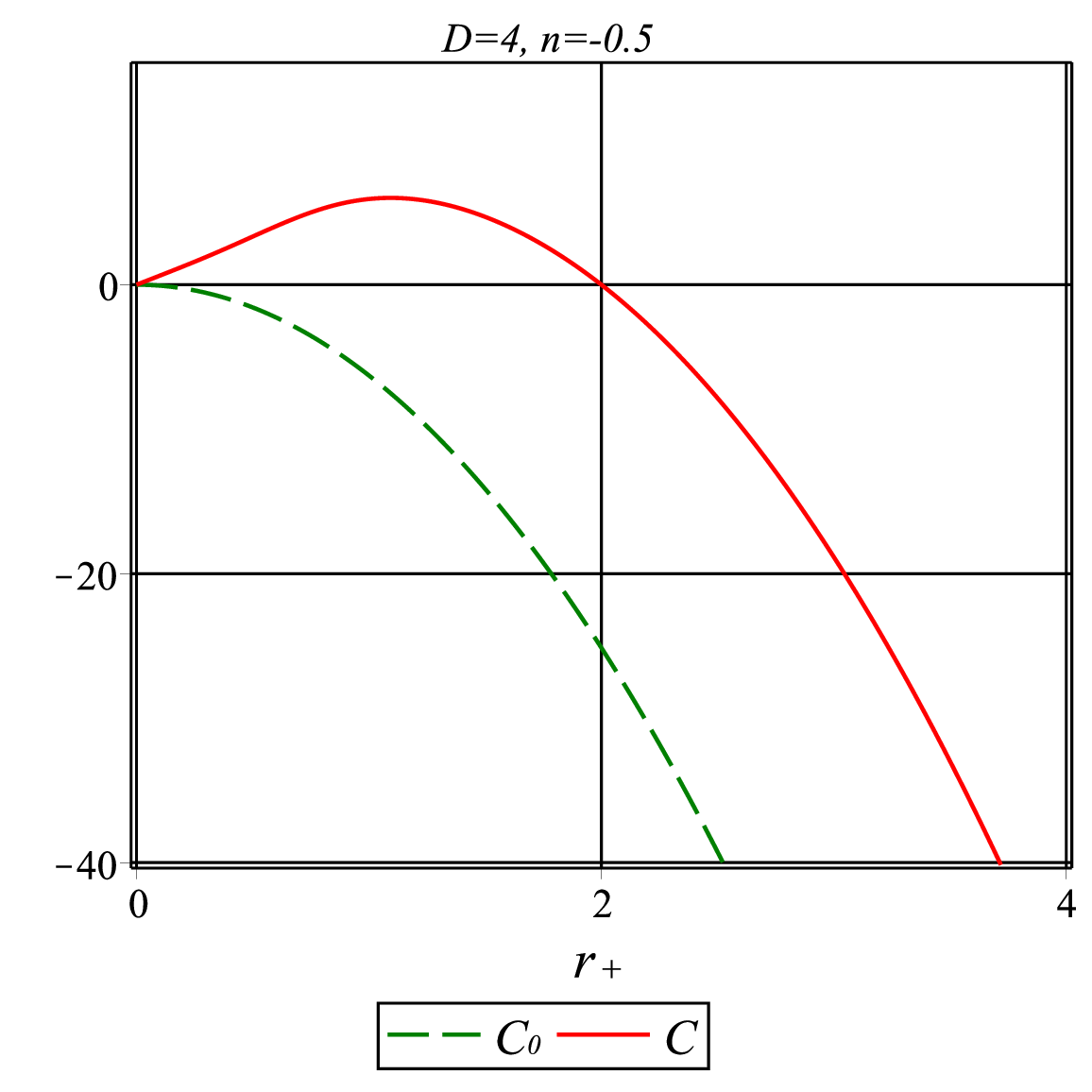}
 \end{array}$
 \end{center}
\caption{Specific heat of $4D$ Schwarzschild black hole in terms of $r_{+}$ for $g=1$.}
 \label{fig8}
\end{figure}

Right plot of Fig. \ref{fig2}, shows the effect of quantum correction on the Schwarzschild black hole pressure. In the case of a large area, the pressure is negative, however, pressure becomes positive at small area due to the thermal fluctuations. In order for seeing a variation of specific heat with the horizon radius, we draw Fig. \ref{fig8}. As we found earlier, the effect of quantum fluctuations may be black hole stability.

\subsection{5D}
\begin{figure}[h!]
 \begin{center}$
 \begin{array}{cccc}
\includegraphics[width=70 mm]{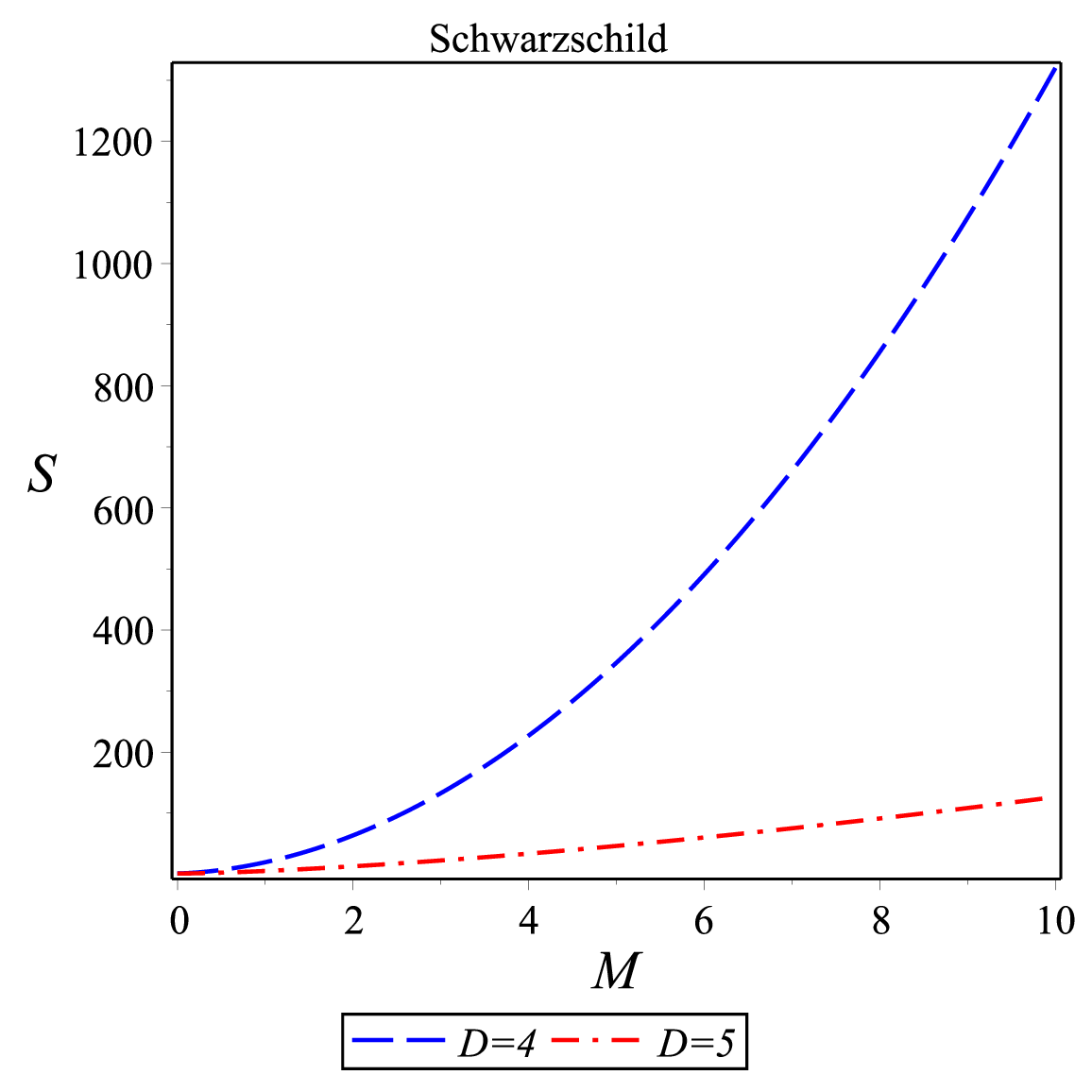}
 \end{array}$
 \end{center}
\caption{Corrected entropy in terms of $M$.}
 \label{fig9}
\end{figure}

Five-dimensional Schwarzschild black hole given by the metric (\ref{metric}) with $D=5$ and (see for example \cite{206.10050}),
\begin{equation}\label{metricS5D}
f(r)=1-\frac{r_{+}^{2}}{r^{2}},
\end{equation}
where $r_{+}^{2}=\frac{8}{3\pi}M_{0}$ is event horizon radius. The black hole entropy and thermodynamic volume is given by,
\begin{equation}\label{SS5D}
S_{0}=\frac{\pi^{2}}{2} r_{+}^{3},
\end{equation}
and
\begin{equation}\label{VS5D}
V=\frac{\pi^{2}}{2} r_{+}^{4},
\end{equation}
respectively.
Using the relations (\ref{temp}) and (\ref{metricS5D}) one can obtain,
\begin{equation}\label{TS5D}
T=\frac{1}{2\pi r_{+}}.
\end{equation}
Combination of (\ref{SS5D}) and (\ref{TS5D}) yields $T=\frac{1}{2(2\pi)^{1/3} S_{0}^{-1/3}}$, hence we find $n=-\frac{1}{3}$ and $c_{1}=\frac{1}{2(2\pi)^{1/3}}$ in the equation (\ref{T}). Also, combination of (\ref{V}), (\ref{SS5D}) and (\ref{VS5D}) gives, $v_{d}=(\frac{2}{\pi^{2}})^{1/3}$, so $V=(\frac{2}{\pi^{2}})^{1/3}S_{0}^{\frac{4}{3}}$.\\
Using the first law of thermodynamics we find
\begin{equation}\label{MS5D}
M=M_{0}-0.24.
\end{equation}
In that case, relation (\ref{SGDS}) hold if the following condition satisfied,
\begin{equation}\label{Smarr-5D}
r_{+}^{4}-\frac{2}{\pi^{2}}r_{+}-\frac{3}{2\pi^{8/3}}=0.
\end{equation}
We can analyze black hole stability by using a sign of specific heat. We can find that specific heat is negative for all area range. Therefore, opposite to the previous case, the $5D$ Schwarzschild black hole is completely unstable even under quantum correction.\\
We can express exponential corrected entropy in terms of corrected mass to see that the second law of thermodynamics is also satisfied which is illustrated by Fig. \ref{fig9}.

\section{Reissner-Nordstr\"{o}m black hole}\label{RN}
\begin{figure}[h!]
 \begin{center}$
 \begin{array}{cccc}
\includegraphics[width=55 mm]{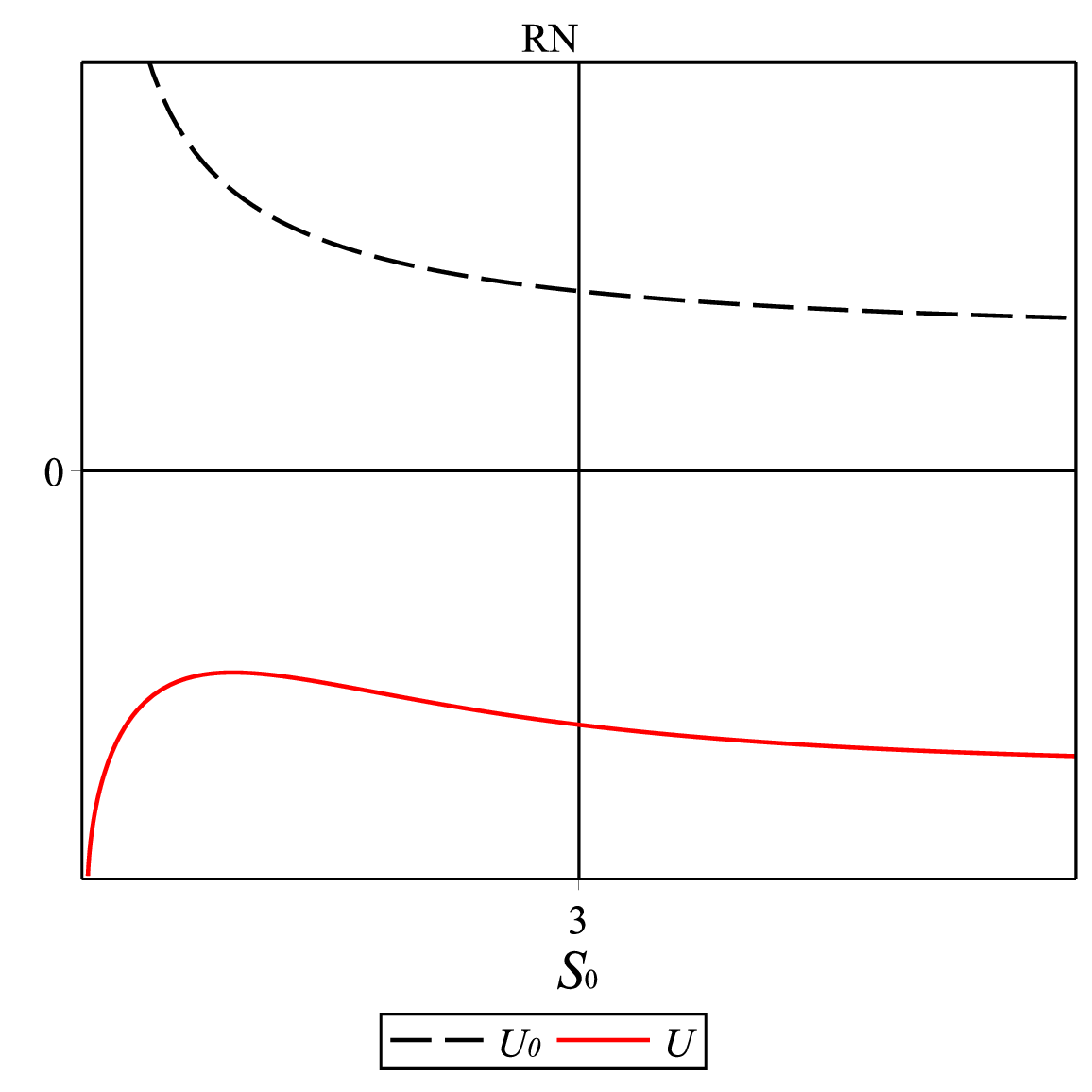}\includegraphics[width=55 mm]{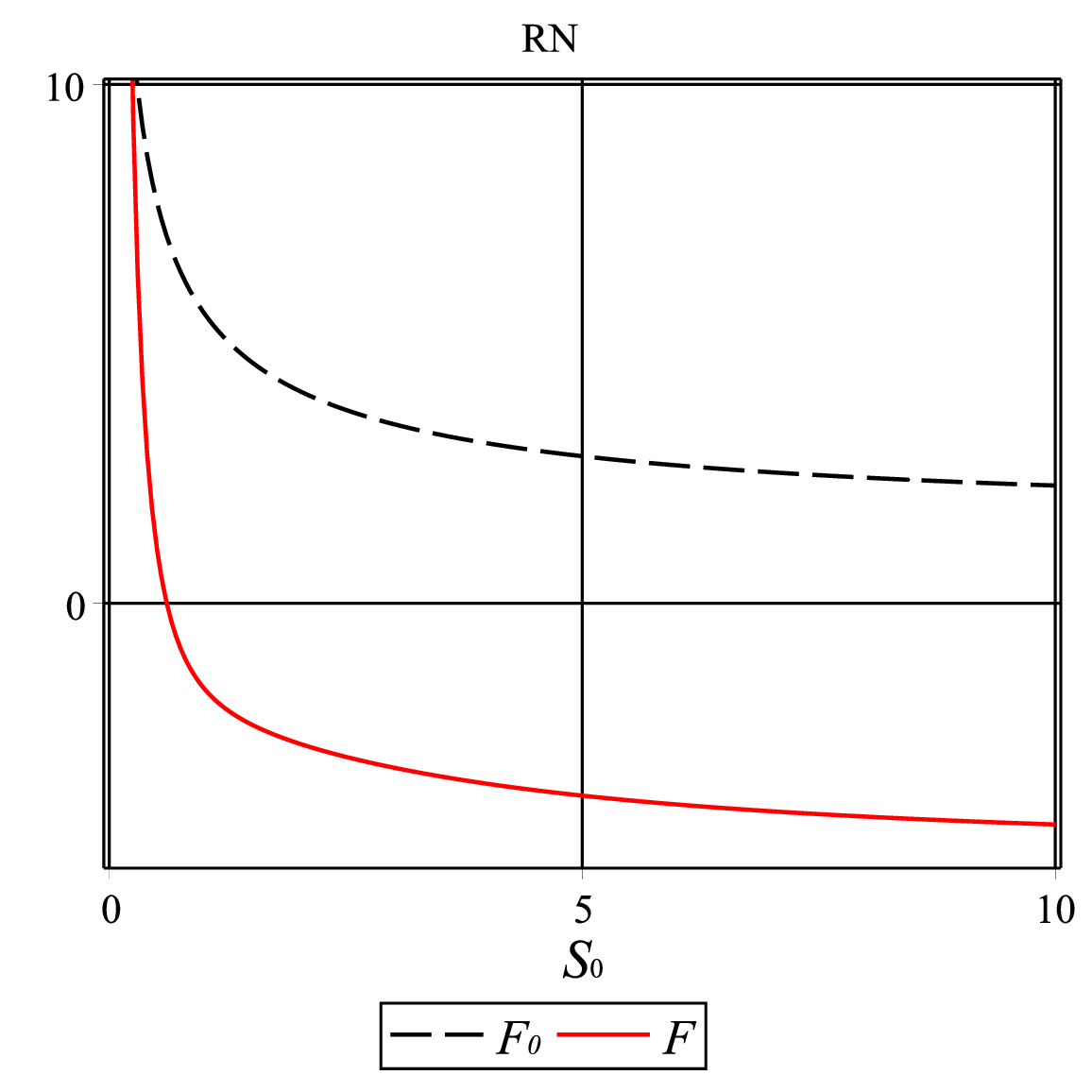}
 \end{array}$
 \end{center}
\caption{Internal and Helmholtz free energies in terms of $S_{0}$ for $Q=g=1$.}
 \label{fig10}
\end{figure}

Now, we consider a charged black hole. Four-dimensional Reissner-Nordstr\"{o}m black hole given by the metric (\ref{metric}) with $D=4$ and (see for example \cite{1401.2586}),
\begin{equation}\label{metricRN}
f(r)=1-\frac{2M_{0}}{r}+\frac{Q^{2}}{r^{2}},
\end{equation}
where $Q$ is the black hole charge. The first law of black hole thermodynamics read as,
\begin{equation}\label{first law RN}
dM=TdS+\Phi dQ,
\end{equation}
where electrostatic potential given by,
\begin{equation}\label{Phi RN}
\Phi=\frac{Q}{r_{+}}.
\end{equation}
It is clear that the first law of black hole thermodynamics satisfied as $4D$ Schwarzschild black hole. Hence, corrected mass is the same as (\ref{MS4D}) and the Smarr-Gibbs-Duhem formula (\ref{SGDS}) should extended to \cite{1401.2586},
\begin{equation}\label{SGDRN}
\frac{D-3}{D-2}M=TS+\frac{D-3}{D-2}\Phi Q,
\end{equation}
which will be hold if the following equation satisfied approximately,
\begin{equation}\label{conditionRN}
2\pi r_{+}^{4}-\pi r_{+}^{3}-2(6\pi Q^{2}+1)r_{+}^{2}+8Q^{2}=0.
\end{equation}
This case described by the equation (\ref{TT}) where we have only two coefficient. Hence, the equation (\ref{T}) extend to
\begin{equation}\label{TRN}
T=c_{1}S_{0}^{n_{1}}+c_{2}S_{0}^{n_{2}},
\end{equation}
where $c_{1}=\frac{1}{4\sqrt{\pi}}$ and $n_{1}=-\frac{1}{2}$ are as $4D$ Schwarzschild black hole, while $c_{2}=-\sqrt{\pi}Q^{2}$ and $n_{2}=-\frac{3}{2}$ are corresponding to Reissner-Nordstr\"{o}m black hole.\\
Therefore, similar to the previous section we can obtain the corrected partition function as,
\begin{equation}\label{Partition RN}
\ln{Z}=\ln{Z_{0}}+\ln{Z_{c}},
\end{equation}
where we defined,
\begin{equation}\label{Partition RN0}
\ln{Z_{0}}=\frac{\left(g+\frac{12\pi Q^{2}+S_{0}}{\sqrt{S_{0}}}\right)S_{0}^{\frac{3}{2}}}{4\pi Q^{2}-S_{0}},
\end{equation}
and
\begin{equation}\label{Partition RNc}
\ln{Z_{c}}=-\left(\frac{12\pi Q^{2}(\frac{2\sqrt{\pi}}{3}+\frac{1}{3\sqrt{S_{0}}e^{S_{0}}}(2-\frac{1}{S_{0}}))+\frac{1}{\sqrt{S_{0}}e^{S_{0}}}+\sqrt{\pi}}{4\pi Q^{2}-S_{0}}\right)S_{0}^{\frac{3}{2}},
\end{equation}
In order to obtain the above solution, we used the following numerical integral,
\begin{equation}\label{intRN}
\int_{0}^{\infty}{e^{-x}dx}=\frac{\sqrt{\pi}}{2}.
\end{equation}
By using (\ref{U}) and (\ref{Fs}) we can obtain the behavior of internal energy and Helmholtz free energy as
\begin{eqnarray}\label{UF RN}
U=U_{0}+U_{c},\nonumber\\
F=F_{0}+F_{c},
\end{eqnarray}
which are illustrated by plots of Fig. \ref{fig10}. Here, $U_{0}$ and $F_{0}$ denote uncorrected energies, while $U_{c}$ and $F_{c}$ are correction terms. We can see that the corrected internal energy is completely negative while the ordinary one is completely positive. Also, we can see that Helmholtz free energy decreased due to quantum corrections. All of these may be a sign of instability in the presence of quantum correction which will be discussed soon by analyzing sign of specific heat.

\begin{figure}[h!]
 \begin{center}$
 \begin{array}{cccc}
\includegraphics[width=70 mm]{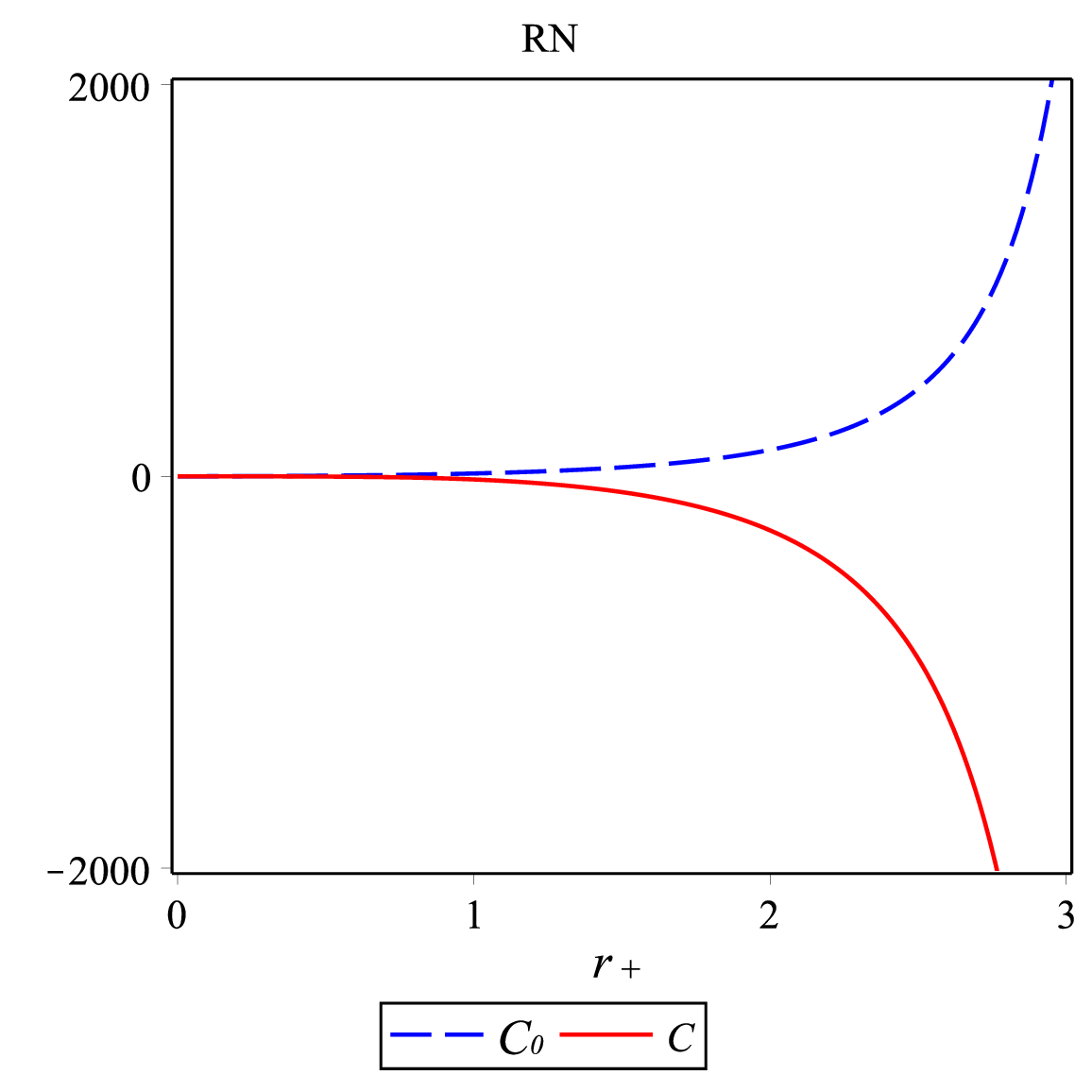}
 \end{array}$
 \end{center}
\caption{Specific heat in terms of $r_{+}$ for $Q=g=1$.}
 \label{fig11}
\end{figure}

A combination of relations (\ref{Cv}) and (\ref{Partition RN}) gives us the specific heat. The presence of $Q$ makes Reissner-Nordstr\"{o}m black hole stable at a small radius while unstable at the large area. So, one can see an unstable/stable phase transition. However, the presence of quantum correction is a cause by instability at a small areas. It is illustrated in Fig. \ref{fig11}.

However, for the charged black holes, we should confirm our stability analysis by using the Hessian matrix of the Helmholtz free energy which is given by,
\begin{eqnarray}\label{mat}
\left(\begin{array}{cc}
\frac{\partial^{2}F}{\partial T^{2}} & \frac{\partial^{2}F}{\partial T\partial \Phi}\\
\frac{\partial^{2}F}{\partial \Phi\partial T}& \frac{\partial^{2}F}{\partial \Phi^{2}}\\
\end{array}\right)=\mathcal{H}.
\end{eqnarray}
It is easy to see that the determinant of the $\mathcal{H}$-matrix vanishes, which means that, one of the eigenvalues is zero. Hence, we should consider the other one which is the trace of the matrix (\ref{mat}) given by,
\begin{equation}\label{trace}
Tr(\mathcal{H})=\left(\frac{\partial^{2}F}{\partial T^{2}}\right)+\left(\frac{\partial^{2}F}{\partial \Phi^{2}}\right).
\end{equation}
Black hole is in stable phase when $Tr(\mathcal{H})\geq0$. Our numerical analysis indicated that $Tr(\mathcal{H})<0$ at a small radius, which confirms black hole instability at quantum scales.

\section{Schwarzschild-AdS black hole}\label{SchAdS}
\begin{figure}[h!]
 \begin{center}$
 \begin{array}{cccc}
\includegraphics[width=70 mm]{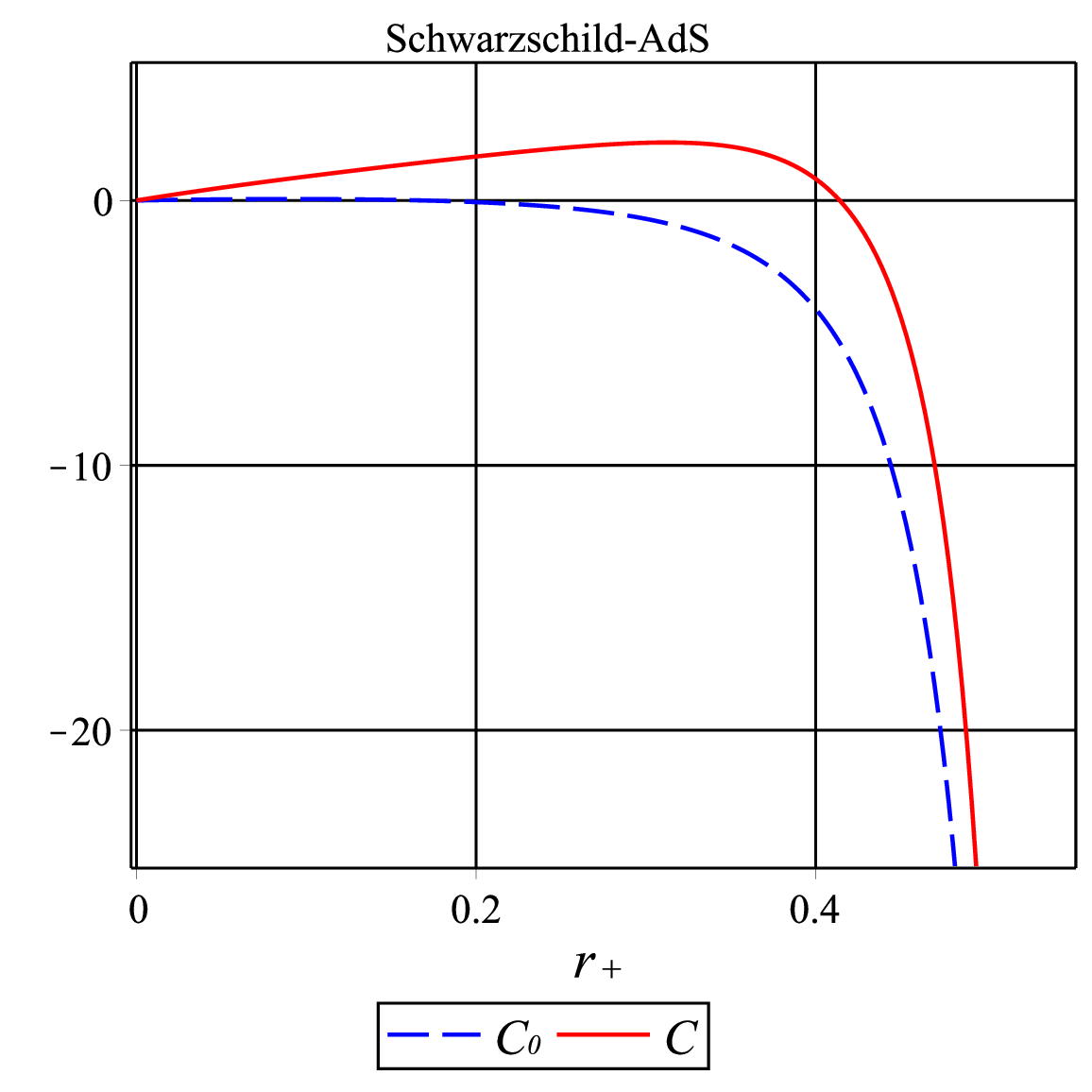}
 \end{array}$
 \end{center}
\caption{Specific heat in terms of $r_{+}$ for $l=g=1$.}
 \label{fig12}
\end{figure}

In this section, we consider the Schwarzschild-AdS black hole in four-dimensional space-time. It is given by the metric (\ref{metric}) with $D=4$ and (see for example \cite{1401.2586}),
\begin{equation}\label{metricSADS}
f(r)=1-\frac{2M_{0}}{r}+\frac{r^{2}}{l^{2}},
\end{equation}
where $l$ is the AdS radius. The first law of black hole thermodynamics read as,
\begin{equation}\label{first law SADS}
dM=TdS+VdP,
\end{equation}
where $V$ is given by the equation (\ref{VS4D}), and thermodynamic pressure is given by,
\begin{equation}\label{P SADS}
P=\frac{3}{8\pi l^{2}}.
\end{equation}
It is clear that the first law of black hole thermodynamics is satisfied. Hence, corrected mass is the same as (\ref{MS4D}) and the Smarr-Gibbs-Duhem formula (\ref{SGDS}) should be extended to \cite{1401.2586},
\begin{equation}\label{SGDSADS}
\frac{D-3}{D-2}M=TS-\frac{2}{D-2}VP,
\end{equation}
which will be held if the following equation satisfied approximately (for $D=4$),
\begin{equation}\label{conditionSADS}
10\pi r_{+}^{4}+2(\pi l^{2}-3)r_{+}^{2}-\pi l^{2}r_{+}-2l^{2}=0.
\end{equation}
Here we have similar relation with the equation (\ref{TRN}) with $c_{1}=\frac{1}{4\sqrt{\pi}}$ and $n_{1}=-\frac{1}{2}$ are as $4D$ Schwarzschild black hole, while $c_{2}=\frac{3}{4\pi^{3/2}l^{2}}$ and $n_{2}=\frac{1}{2}$ are corresponding to Schwarzschild-AdS black hole.\\
The corrected partition function is obtained as following,
\begin{equation}\label{PartitionSADS}
Z=\exp\left(\frac{S_{0}^{2}-\pi l^{2}S_{0}+g\sqrt{S_{0}}}{\pi l^{2}+3S_{0}}\right)\exp\left(\frac{\sqrt{S_{0}}}{\pi l^{2}+3S_{0}}
\left[\pi l^{2}(\sqrt{\pi}+\frac{e^{-S_{0}}}{\sqrt{S_{0}}})+\frac{3\sqrt{\pi}}{2}\right]\right),
\end{equation}
where the last exponential is due to the quantum correction. We find that increasing partition function is a consequence of quantum corrections.\\
Using the equation (\ref{Cv}) we can obtain specific heat and find that quantum corrections make Schwarzschild-AdS black hole stable at the small areas (see Fig. \ref{fig12}). In absence of the thermal fluctuations, Schwarzschild-AdS black hole is unstable at a small horizon radius, while in the presence of quantum correction it will be stable by reducing size.

\section{Charged AdS black hole}\label{Ch}
\begin{figure}[h!]
 \begin{center}$
 \begin{array}{cccc}
\includegraphics[width=70 mm]{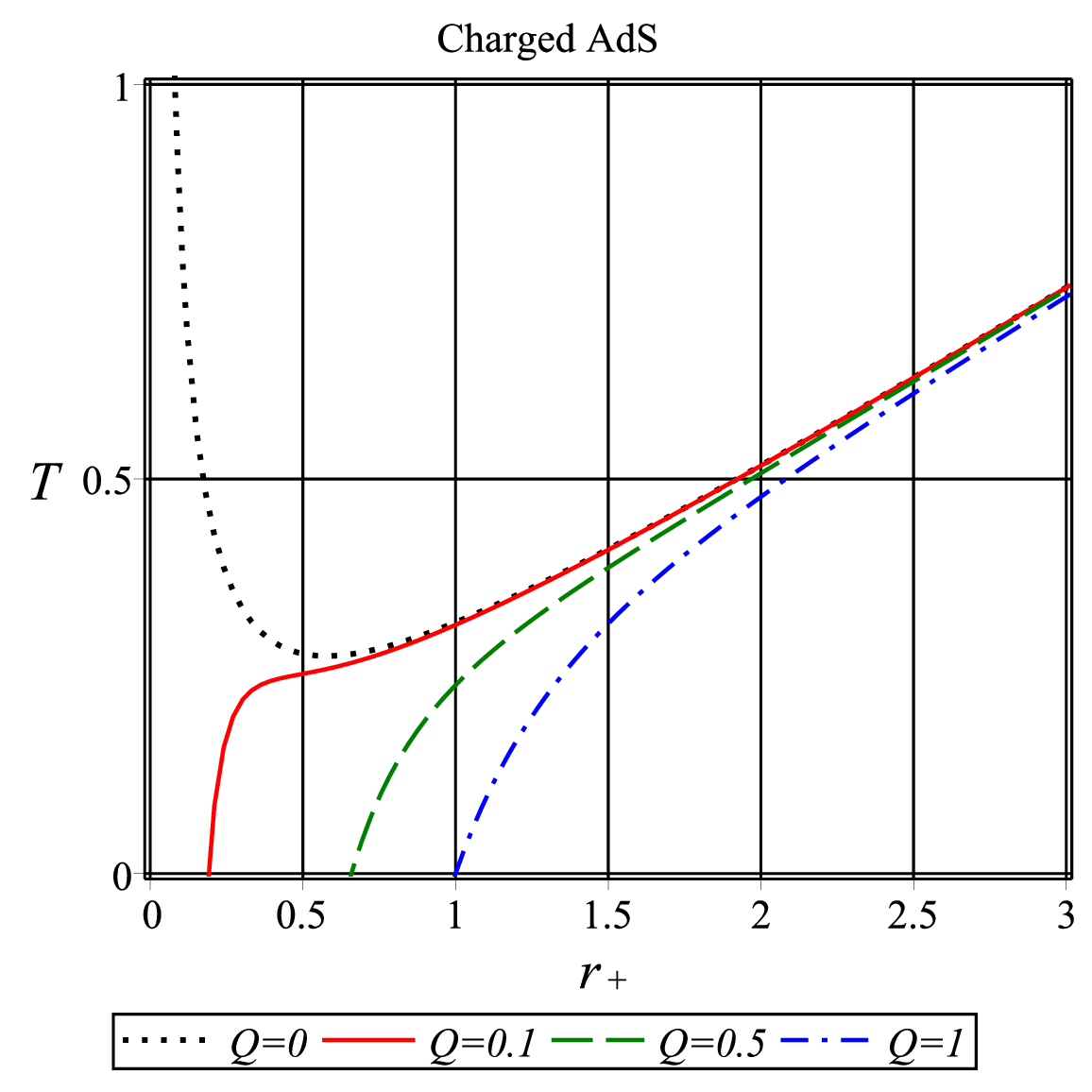}
 \end{array}$
 \end{center}
\caption{Temperature in terms of $r_{+}$ for $l=1$.}
 \label{fig13}
\end{figure}

It is the most general case which considers in this paper. The charged AdS black hole is described by mass, charge, and AdS radius which all of them introduced already. It is given by the metric (\ref{metric}) with $D=4$ and (see for example \cite{1401.2586}),
\begin{equation}\label{metricMQl}
f(r)=1-\frac{2M_{0}}{r}+\frac{Q^{2}}{r^{2}}+\frac{r^{2}}{l^{2}}.
\end{equation}
The first law of black hole thermodynamics read as,
\begin{equation}\label{first law MQL}
dM=TdS+\Phi dQ+VdP,
\end{equation}
where $S$ is given by (\ref{SS4D}), $\Phi$ is given by (\ref{Phi RN}), $V$ is given by the equation (\ref{VS4D}), and pressure $P$ is given by (\ref{P SADS}).
Since the exponential corrected entropy is independent of $Q$ and $P$, it is clear that the first law of black hole thermodynamics is satisfied with the corrected mass (\ref{MS4D}). Hence, the Smarr-Gibbs-Duhem formula given by \cite{1401.2586},
\begin{equation}\label{SGDMQL}
\frac{D-3}{D-2}M=TS+\frac{D-3}{D-2}\Phi Q-\frac{2}{D-2}VP.
\end{equation}
Using the corrected mass and entropy for $D=4$ we find that the Smarr-Gibbs-Duhem formula satisfied if,
\begin{equation}\label{conditionMQL}
4\pi r_{+}^{6}+2\pi (l^{2}+3)r_{+}^{4}-\pi l^{2}r_{+}^{3}-2(3\pi Q^{2}l^{2}+ l^{2}+3)r_{+}^{2}+2Q^{2}l^{2}=0.
\end{equation}
Here, we have the equation (\ref{T}) with three terms as,
\begin{equation}\label{TMQL}
T=\frac{1}{4\sqrt{\pi}}S_{0}^{-\frac{1}{2}}-\sqrt{\pi}Q^{2}S_{0}^{-\frac{3}{2}}+\frac{3}{4\pi^{3/2}l^{2}}S_{0}^{\frac{1}{2}}.
\end{equation}
In Fig. \ref{fig13} we can see typical behavior of the black hole temperature. In the case of charged AdS black hole, we can see a minimum horizon radius where temperature vanishes. Hence, it is interesting to find what happens for the black hole in the presence of quantum corrections where its temperature becomes zero.\\

\begin{figure}[h!]
 \begin{center}$
 \begin{array}{cccc}
\includegraphics[width=70 mm]{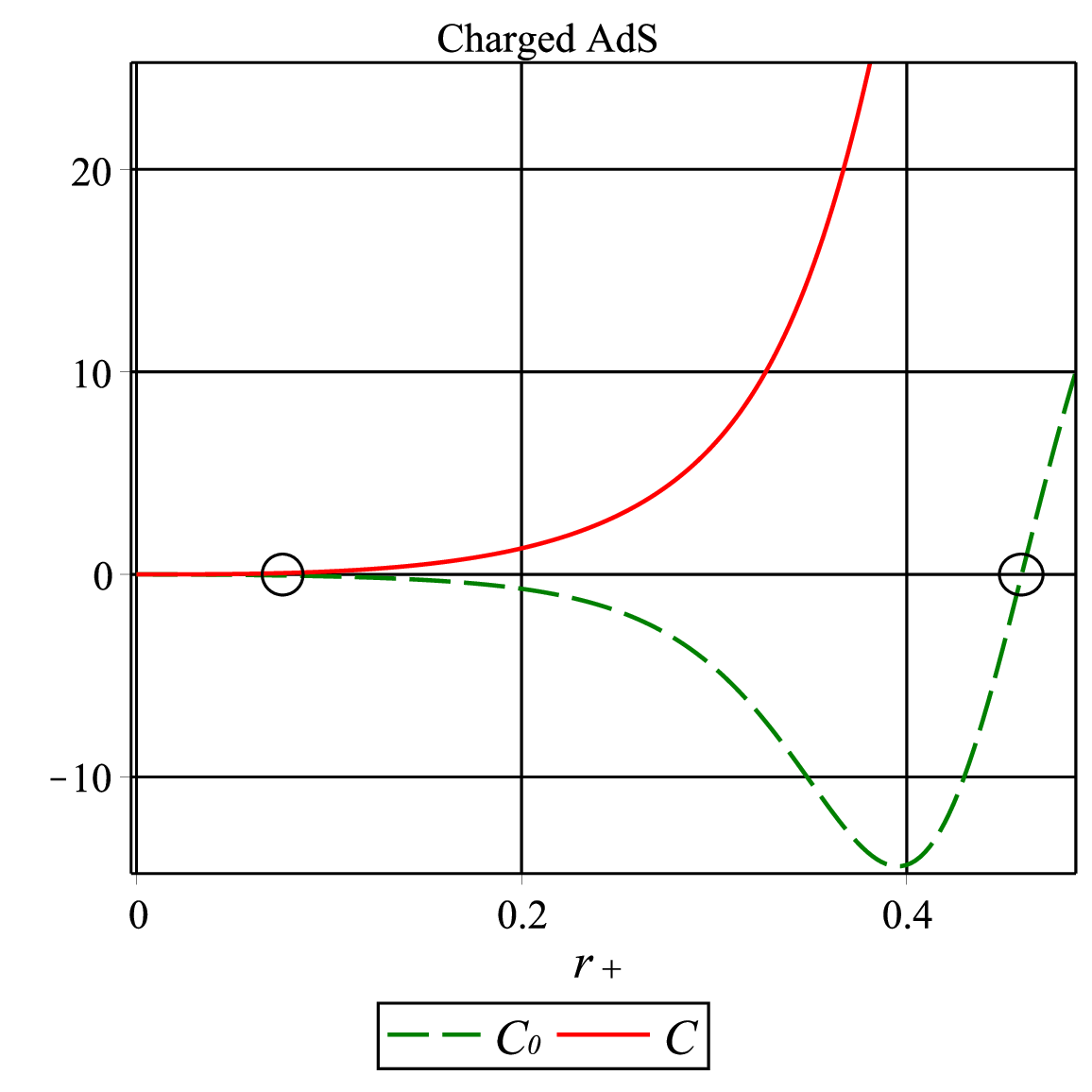}
 \end{array}$
 \end{center}
\caption{Specific heat in terms of $r_{+}$ for $l=g=1$, and $Q=0.1$.}
 \label{fig14}
\end{figure}

In that case, corrected partition function obtained similar to (\ref{Partition RN}) where,
\begin{equation}\label{Z0QML}
\ln{Z_{0}}=\frac{(g-3\sqrt{\pi})\sqrt{S_{0}}+24\pi^{2}Q^{2}l^{2}+2\pi l^{2}S_{0}-2S_{0}^{2}}{8\pi^{2}Q^{2}l^{2}-2\pi l^{2}S_{0}-6S_{0}^{2}}S_{0},
\end{equation}
is ordinary partition function and
\begin{equation}\label{ZcQML}
\ln{Z_{c}}=-\frac{\pi l^{2}\left(\sqrt{\pi}S_{0}^{3/2}e^{S_{0}}(8\pi Q^{2}+1)+4\pi Q^{2}(2S_{0}-1)+S_{0}\right)}{(8\pi^{2}Q^{2}l^{2}-2\pi l^{2}S_{0}-6S_{0}^{2})e^{S_{0}}},
\end{equation}
is corrected terms, hence corrected partition function is $\ln{Z}=\ln{Z_{0}}+\ln{Z_{c}}$.\\
Now, we can study specific heat graphically. By using the above relations in the equation (\ref{Cv}) we can find that the specific heat of charged AdS black hole is negative at a small radius while in the presence of quantum correction, specific heat is positive. These are illustrated in Fig. \ref{fig14}. The dashed green line of Fig. \ref{fig14} shows that charged AdS black hole is stable when $r_{1}<r_{+}<r_{2}$ in agreement with the claim of Ref. \cite{1401.2586} ($r_{1}$ and $r_{2}$ denoted by circles in Fig. \ref{fig14}). We can confirm such stability by analyzing the Hessian matrix of the Helmholtz free energy similar to the section \ref{RN}.\\

\begin{figure}[h!]
 \begin{center}$
 \begin{array}{cccc}
\includegraphics[width=70 mm]{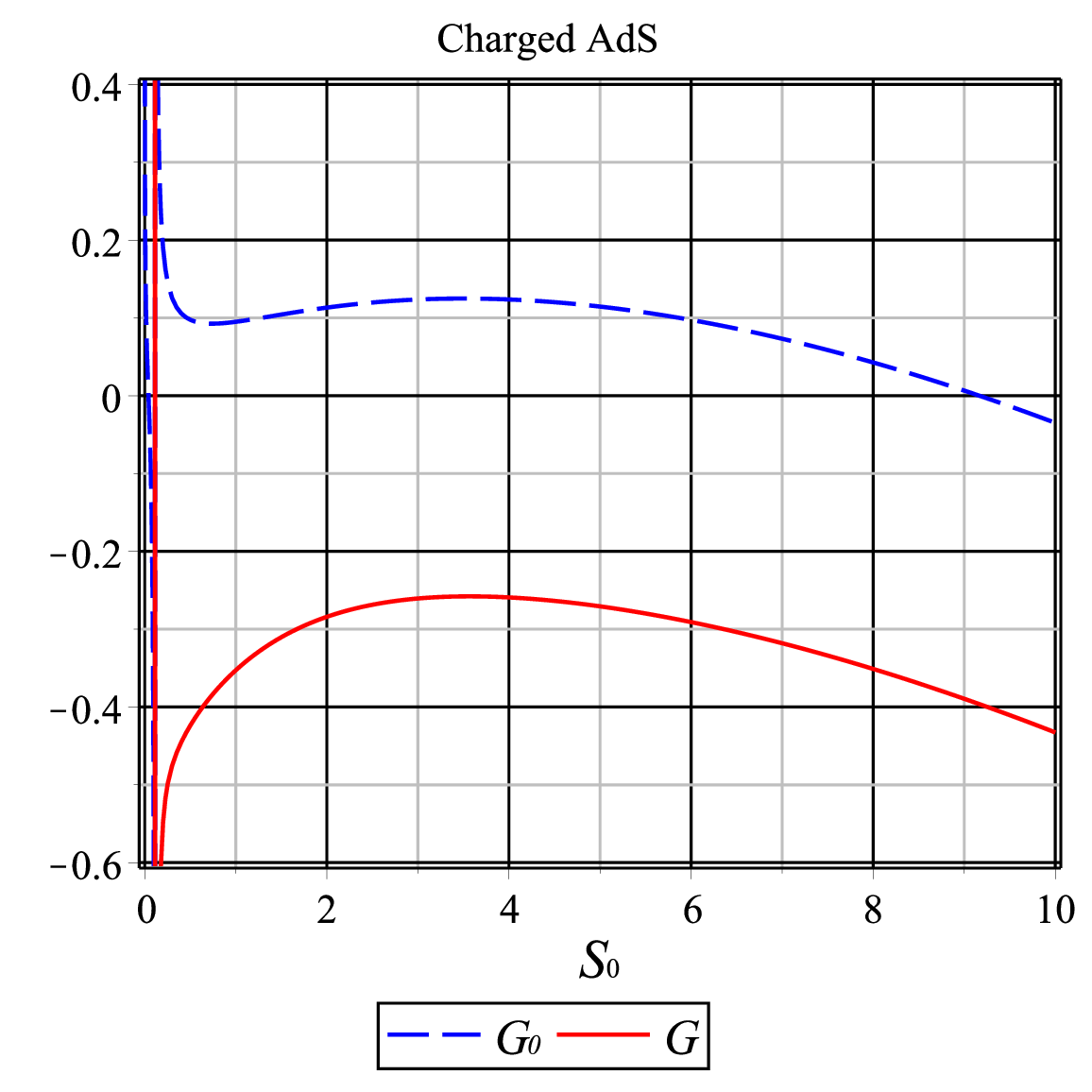}
 \end{array}$
 \end{center}
\caption{Gibbs free energy in terms of $S_{0}$ for $l=g=1$, and $Q=0.1$.}
 \label{fig15}
\end{figure}

By using the result of Fig. \ref{fig13} we can see that the black hole temperature is zero at about $r_{+}=0.2$ (for $Q=0.1$), however, at this point, we have non-zero mass, entropy, and specific heat. We can interpret it as a black remnant.\\
Then we can see the typical behavior of Gibbs free energy in Fig. \ref{fig15}. We can see some local extremum in Gibbs free energy, anyway effect of quantum correction is decreasing it. We find that opposite to the ordinary case \cite{1401.2586} there is no holographically dual Van der Waals fluid in presence of quantum corrections (in small area), hence there is no any critical points and phase transition at quantum scales.\\

\section{Conclusion}\label{Con}
In this paper, we considered the recent idea that the black hole entropy should be corrected by exponential term \cite{2007.15401} at quantum scales, and shown that such corrected entropy obtained by modified black hole metric. The main goal of this paper was study of such an effect on the black hole thermodynamics. Hence, the effect of the exponential correction term of the entropy on the canonical partition function is calculated. Therefore, corrected partition function due to the quantum fluctuations obtained which is used to study exponential corrected thermodynamics of some black holes. We found that such a correction term affects the black hole stability when the black hole size becomes small.
We found that as a $4D$ Schwarzschild black hole, which decreased its size due to the Hawking radiation, becomes stable at a small area limit hence does not evaporate. We obtained a similar result for the Schwarzschild-AdS black hole in four-dimensional space-time. On the other hand, we found that the quantum corrections have no effect on the $5D$ Schwarzschild black hole. Opposite to the $4D$ Schwarzschild black hole, we found that Reissner-Nordstr\"{o}m black hole (charged black hole) is unstable due to thermal fluctuations. It means that the black hole charge is a cause by a stable/unstable phase transition which happen when the black hole size is large. However, the charged AdS black hole will be stable at a small area due to the quantum correction, which may yield to the black remnant. It means that the stable black hole evaporates and creates the black remnants at zero temperature while non-zero mass. It may be used to solve the information loss paradox of black holes.\\
For the future work, it is interesting to find the exponential corrected partition function of Kerr or Kerr-AdS black holes (rotating black holes \cite{1401.2586}) and discuss the quantum effect on thermodynamics stability.  Also, it would be interesting to apply method used in this paper for BTZ \cite{jac} or Ho\v{r}ava-Lifshitz black holes \cite{EPJC003}.\\
Already, we used logarithmic corrected entropy \cite{QGP} to study shear viscosity to entropy ratio and found the universal lower bound violated. It is interesting to consider exponential corrected entropy to study the holographic description of a quark-gluon plasma \cite{QCD}.

\section*{Acknowledgments}

B.P. would like to thank Iran Science Elites Federation, Tehran,
Iran, and Canadian Quantum Research Center 204-3002 32 Ave Vernon, BC V1T 2L7 Canada.

\end{document}